\documentclass[journal]{IEEEtran}
\usepackage{amsmath,amsfonts}
\usepackage{array}
\usepackage{textcomp}
\usepackage{stfloats}
\usepackage{cite}
\usepackage{amsmath,amssymb,amsfonts,bm}
\usepackage{amsthm}
\usepackage{graphicx}
\usepackage{subcaption}
\usepackage{textcomp}
\theoremstyle{remark}
\usepackage{algorithm}
\usepackage{algpseudocode} 
\newtheorem{remark}{Remark}
\usepackage{xcolor}
\usepackage{tikz}
\usepackage{setspace}
\usepackage{titlesec}
\usepackage{graphicx}
\usepackage{caption} 
\usepackage{array}
\usepackage{booktabs}
\usepackage{mathrsfs}
\usepackage{booktabs}
\usepackage{tabularx}
\usepackage{url}
\usepackage{verbatim}
\usepackage{graphicx}
\usepackage{cite}
    \title{{Delay-Doppler Domain Aided OFDM (DD-a-OFDM): Pulsone Pilot and Cross-Domain Iterative Receiver Design}}
    
	\author{Rongrong Xin, \IEEEmembership{Student Member, IEEE}, Zhiqiang Wei, \IEEEmembership{Member, IEEE}, Zai Yang, \IEEEmembership{Senior Member, IEEE} \thanks{R. Xin, Z. Wei, Z. Yang are with Xi'an Jiaotong University, Xi'an 710049, China (e-mail: Xinrongrong@stu.xjtu.edu.cn; zhiqiang.wei@xjtu.edu.cn; yangzai@xjtu.edu.cn). } \vspace{-8mm}}

\begin{document}

	\maketitle
	\begin{abstract}
    In high-mobility channels, the widely adopted orthogonal frequency division multiplexing (OFDM) modulation waveform suffers from significant performance degradation due to channel variation and severe inter-carrier interference. Although delay–Doppler (DD) domain multicarrier (DDMC) modulation enjoys a higher channel estimation (CE) accuracy via exploiting the inherent DD-domain channel stability, sparsity, and separability, its high detection complexity limits its practical deployment, making OFDM still a promising waveform candidate for the sixth-generation (6G) cellular networks. To simultaneously harness the complementary strengths of DDMC and OFDM and mitigate their respective limitations, this paper proposes a DD-domain aided OFDM (DD-a-OFDM) framework, including a pulsone pilot (PP) design and a cross-domain iterative receiver design, to enhance the CE and detection performance of OFDM in high-mobility channels. Specifically, a pulsone pilot sequence is periodically inserted before the cyclic prefix (CP) for each OFDM symbol in the time-domain, which is equivalent to a single DD-domain pilot impulse and enables accurate DD-domain CE. Then, we propose a cross-domain iterative receiver in which the time-frequency (TF) domain detection results are transformed into the DD-domain for interference cancellation and CE; in turn, the DD-domain estimated channel is transformed into the TF-domain for channel reconstruction and symbol detection. Owing to the DD-domain channel sparsity, the coherent combination of PP symbols from the time-domain to the DD-domain, as well as the incoherent combination of data symbols from the TF-domain to the DD-domain, the proposed cross-domain iterative receiver achieves a higher CE accuracy and detection performance with a much lower pilot overhead, compared to the conventional OFDM system. Simulation results validate the fast convergence and superior performance of the proposed DD-a-OFDM scheme and demonstrate that it maintains a robust CE and detection performance even in high-mobility channels.
	\end{abstract}
	\vspace{-1mm}
    \vspace{-3mm}
	\section{INTRODUCTION}
    \vspace{-1mm}
    \IEEEPARstart{I}{n} the sixth-generation (6G) era, high-mobility communication scenarios are becoming increasingly prevalent, with representative applications including high-speed railway communications \cite{Railways}, low-altitude wireless networks \cite{low_altitude,UAV}, and satellite communications \cite{Satellite}. Currently, the fifth-generation (5G) cellular network relies primarily on orthogonal frequency division multiplexing (OFDM) as its core modulation scheme. However, in high-mobility environments, rapidly time-varying channels necessitate frequent channel estimation (CE) and disrupt subcarrier orthogonality in OFDM systems, leading to significant performance degradation. Against this backdrop, delay–Doppler (DD) domain multicarrier (DDMC) modulation has attracted considerable attention due to its ability to exploit the inherent stability, sparsity, and separability of DD-domain channels\cite{OTFS_1_Hadani,Shen,OTFS_SHUANGYANG,OTFS_wei,ODDM,Diversity_1,Diversity_Li}. Although DDMC exhibits superior CE and data detection performance in high-mobility channels, it suffers from a higher detection complexity compared to OFDM.

    A natural question arises: how can the complementary strengths of DDMC and OFDM be jointly leveraged while mitigating their respective limitations? To address this, this paper proposes a DD-domain aided OFDM (DD-a-OFDM) transceiver architecture that integrates the precise CE capability of DDMC with the low-complexity equalization of OFDM, delivering robust performance in high-mobility environments.
    
    The widespread adoption of OFDM largely relies on several key benefits. First, the insertion of a cyclic prefix (CP) converts linear convolution into circular convolution, allowing the channel matrix to be diagonalized by the Fourier transform, which reduces equalization to a set of simple one-tap per-subcarrier operations. Second, the orthogonal subcarrier structure enables a high spectral efficiency by densely packing subcarriers without inter-carrier interference (ICI). Third, OFDM offers flexible resource allocation, supporting adaptive modulation and coding, frequency-domain scheduling, and scalable bandwidth configurations. However, in high-mobility channels, significant Doppler spread induces distinct frequency shifts across subcarriers, thereby destroying their orthogonality and leading to severe ICI. Moreover, CE schemes in OFDM systems typically rely on pilots inserted in the TF-domain. The channel coefficients at the pilot TF bins are first estimated using the least-squares (LS) or minimum mean square error (MMSE) estimator, and the full channel response is then reconstructed via interpolation across TF grids \cite{COFDM2}. In doubly-selective fading channels, the pilot density should be increased in both the time and frequency dimensions to achieve an acceptable CE accuracy. Such pilot arrangements inevitably consume substantial TF resources, thereby reducing the effective spectral efficiency.

    DDMC techniques have recently attracted significant research attention. Representative schemes include orthogonal time–frequency space (OTFS) \cite{OTFS_1_Hadani,OTFS_wei,OTFS_SHUANGYANG,OTFS_NIE}, orthogonal delay–Doppler division multiplexing (ODDM) \cite{ODDM}, and delay–Doppler orthogonal pulse (DDOP) \cite{DDOP}. By modulating symbols in the DD-domain and spreading into the entire TF-domain, DDMC has the potential to exploit the full diversity gain of the TF-domain channel\cite{Diversity_1,Diversity_Li}, thereby demonstrating greater robustness and reliability than conventional OFDM, particularly in high-mobility scenarios. More importantly, DDMC enables the transformation of the TF-domain doubly-selective fading channel into the DD-domain quasi-static and sparse channel, providing a solid foundation for low-complexity and high-accuracy CE schemes. A variety of CE approaches have been proposed in the DD-domain, including simple threshold-based methods \cite{Thre2}, advanced compressed sensing (CS)- based approaches \cite{off_grid}, superimposed pilot-based schemes \cite{yuan_data_aided}, and neural network-based approaches \cite{DL_OTFS1,DL_OTFS3}, which collectively enrich the existing DD-domain CE framework.
  
    Given the well-established deployment of OFDM and the practical challenges of fully replacing it in high-mobility scenarios, OFDM remains one of the most promising waveform candidates for 6G systems \cite{OFDM_6G}. Therefore, DD-a-OFDM is an important research direction for effectively exploiting the advantages of the DD-domain channel and signal processing, while maintaining compatibility with existing 5G OFDM systems, thereby enabling the smooth evolution of current communication networks. Nevertheless, only a limited number of studies have investigated this issue. For instance, in \cite{Nie}, the authors proposed a cross-domain CE algorithm applicable to conventional OFDM systems. In this approach, the received TF-domain pilot sequences of OFDM are transformed into the DD-domain, where a two-dimensional (2D) twisted convolution is performed to obtain coarse estimates of the channel delay and Doppler parameters. Subsequently, based on a dictionary constructed from the estimated delay and Doppler components, the OFDM CE problem is reformulated as a sparse signal recovery (SSR) problem in the TF-domain and solved via the famous least absolute shrinkage and selection operator (LASSO). In \cite{DDaOFDM}, the authors proposed a DD-a-OFDM scheme, which preserves the conventional OFDM transceiver structure and places pilots regularly on the TF grid. Then, maximum likelihood (ML) and peak detection-based channel estimators, along with the corresponding TF-domain equalizer, are developed to enhance OFDM performance in high-mobility scenarios. However, these studies rely solely on the TF-domain pilot design and fail to fully exploit the DD-domain channel property. Moreover, the estimation accuracy of delay and Doppler parameters is constrained by the number of pilots in the TF-domain and the spectral efficiency still leaves room for further improvement.

    In the context of DD-a-OFDM, this paper proposes a novel pilot design and a cross-domain transceiver architecture that is compatible with OFDM\cite{5GNR1,5GNR2,5GNR3}, and exploits the DD-domain channel property to mitigate performance degradation of OFDM in high-mobility channels. The main contributions of this paper are summarized as follows:
    \begin{itemize}
        \item This paper proposes a novel DD-a-OFDM transceiver architecture, in which pilots are innovatively placed in the DD-domain while data symbols reside in the TF-domain, which has the following features: (1) a single DD-domain pulse corresponds to a periodic pulse train in the time-domain, which is referred to as pulsone pilot (PP) sequence in this paper, allowing the OFDM symbol to be naturally embedded among them, thus achieving inherent compatibility with OFDM; (2) \textcolor{black}{conventional OFDM and DD-a-OFDM CE methods exploit limited TF-domain resources for pilot transmission, which limits the achievable delay and Doppler estimation accuracy \cite{COFDM2,DDaOFDM,Nie}. In contrast, the proposed cross-domain pilot design enables a single DD-domain pilot to leverage the entire TF-domain resources, achieving high-resolution delay-Doppler estimation with a low pilot overhead;} (3) the cross-domain pilot design can fully exploit the stability, sparsity and separability of the DD-domain channel, enabling high-precision CE.
        \item A cross-domain receiving signal processing and interference analysis framework is proposed, where received signals are transformed into the TF-domain for symbol detection and into the DD-domain for CE, yielding the following key insights: (1) due to the CP in OFDM systems, the received TF-domain signals are free from interference caused by the DD-domain pilots; (2) in the DD-domain, the received time-domain PP sequence undergoes coherent combining, while the received TF domain data symbols are incoherently combined and behave approximately as Gaussian noise, \textcolor{black}{thereby enhancing the pilot-to-interference power ratio in the DD-domain observation and improving CE performance.}
        \item A \textbf{C}ross-\textbf{D}omain iterative \textbf{I}nterference cancellation (IC), channel \textbf{E}stimation and symbol \textbf{D}etection framework is then developed, which we refer to as CD-IED in this work. Specifically, the DD-domain interference is reconstructed and canceled based on the symbol detection and CE results from the previous iteration, thereby reducing the DD-domain interference level, enhancing CE accuracy, and further improving the TF-domain symbol detection performance. Through iterative cross-domain information exchange and joint processing, both CE accuracy and symbol detection performance are progressively improved.
        \item The performance of the proposed algorithm is validated through numerical simulations. Simulation results show that the iterative algorithm typically converges rapidly within 3 iterations, exhibiting stable convergence behavior. More importantly, the proposed framework demonstrates strong robustness against Doppler effects and consistently outperforms the conventional OFDM transceiver architecture.
    \end{itemize}
    $\textit{Notations:}$ Bold face lower case and upper case letters represent vector and matrix, respectively. The blackboard lettters $\mathbb{A}$, $\mathbb{C}$ denote the constellation set and complex number filed. $\textbf{I}_{N}$ denotes the identity matrix of size  $\textit{N}\times\textit{N}$. $\textbf{0}_{N}$ denotes the zero matrix of size  $\textit{N}\times\textit{N}$. $\otimes$ denotes the Kronecker product operator. $(\cdot)^\mathrm{H}$ denotes the Hermitian transpose. $\overline{(\cdot)}$ and ${(\cdot)}^*$ represents the complex conjugate. $(\cdot)^\mathrm{T}$ denotes the transpose. vec$(\cdot)$ denotes the vectorized version of a matrix by stacking the elements column by column, while vec$^{-1}$$(\cdot)$ denotes the corresponding inverse process. diag$(\cdot)$ denotes the diagonal matrix. $\textbf{F}_{N}$  and  $\textbf{F}_{N}^{H} $ denote the discrete Fourier transform (DFT) matrix and inverse DFT (IDFT) of size  $\textit{N}\times\textit{N}$. $[\cdot]_M$ denotes mod-$M$ operation.

    \vspace{-2mm}
	\section{\textcolor{black}{The proposed Transceiver Architecture and System Model}}
	\label{SYSTEM MODEL}

    \textcolor{black}{This section introduces the proposed cross-domain pilot design and the corresponding transceiver architecture.}

    \vspace{-2mm}
    \subsection{DD-a-OFDM Transceiver Architecture}
\begin{figure}[t]
    \centering
    \vspace{-3mm}        \includegraphics[width=0.9\linewidth]{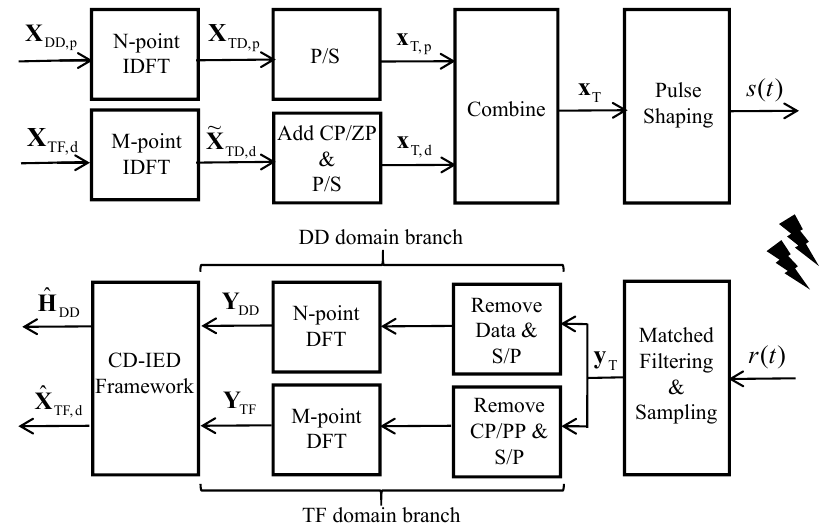}\vspace{-2mm}
    \caption{\textcolor{black}{Proposed DD-a-OFDM transceiver architecture.}}
            \label{Transceiver}
    \end{figure}

   \textcolor{black}{Before delving into the details, we present an overview of the proposed DD-a-OFDM transceiver architecture, as illustrated in Fig. \ref{Transceiver}. Let \(N\) and \(M\) denote the number of OFDM symbols and subcarriers, respectively. At the transmitter, the data symbols $\mathbf{X}_{\mathrm{TF,d}}$ are mapped onto the TF grid. Then, an $M$-point IDFT is applied along the frequency dimension, followed by the insertion of a cyclic-prefix (CP) and zero-padding (ZP). After parallel-to-serial (P/S) conversion, time-domain discrete data signals $\mathbf{x}_{\mathrm{T,d}}$ are obtained. In contrast to conventional designs, the pilot symbols $\mathbf{X}_{\mathrm{DD,p}}$ are placed in the DD-domain. Through an $N$-point IDFT along the Doppler dimension and P/S conversion, the corresponding time-domain pilot signals $\mathbf{x}_{\mathrm{T,p}}$ are obtained. Finally, $\mathbf{x}_{\mathrm{T,d}}$ and $\mathbf{x}_{\mathrm{T,p}}$ are combined to form the overall time-domain transmit signal $\mathbf{x}_{\mathrm{T}}$, which is subsequently transmitted after pulse shaping.}

    \textcolor{black}{At the receiver, the received signal $r(t)$ is first processed by matched filtering and sampling to obtain the discrete time-domain received signal $\mathbf{y}_{\mathrm{T}}$. It is then processed through two parallel branches and transformed into the TF and DD-domains for symbol detection and CE, respectively. In the TF-domain branch, the CP and PP are removed, followed by serial-to-parallel (S/P) conversion and an $M$-point DFT, yielding the received TF-domain signals $\mathbf{Y}_{\mathrm{TF}}$ for data detection. In the DD-domain branch, the time-domain samples that are exclusively contributed by TF-domain data symbols are discarded, whereas those containing pilot components are preserved, which are then transformed back to the DD-domain via an $N$-point DFT, yielding $\mathbf{Y}_{\mathrm{DD}}$ for CE. It is worth noting that TF-domain data symbols interfere with the pilot symbols during transmission, leading to DD-domain interference and degrading CE accuracy. To address this issue, a low-complexity cross-domain iterative receiver framework, termed CD-IED, is proposed, which iteratively performs IC, CE and symbol detection, and finally outputs the refined channel estimates $\hat{\mathbf{H}}_{\mathrm{DD}}$ and detected data symbols $\hat{\mathbf{X}}_{\mathrm{TF,d}}$.}

    \vspace{-3mm}
    \subsection{\textcolor{black}{TF-domain Data Symbol and DD-domain Pilot Symbol Multiplexing Scheme}}

    Define the OFDM symbol duration and subcarrier spacing as \(MT_{s}\) and \(\Delta f\), with $T_{s}=\frac{1}{M\Delta f}$. The transmitted data symbols in the TF-domain \(\mathbf{X}_{\mathrm{TF,d}} \in \mathbb{C}^{M \times N}\) are indexed by \(m \in \{0, 1, \ldots, M-1\}\), \(n \in \{0, 1, \ldots, N-1\}\) and each symbol $\mathbf{X}_{\mathrm{TF,d}}[m,n]$ is drawn from the constellation set $\mathcal{A}$. After applying an $M$-point IDFT in the frequency domain, the resulting time-delay (TD) domain signals are given by
    \vspace{-2mm}
    \begin{equation}\textcolor{black}{
    \tilde{\mathbf{X}}_{\mathrm{TD,d}}[l,n] = \frac{1}{\sqrt{M}} \sum\nolimits_{m=0}^{M-1}\mathbf{X}_{\mathrm{TF,d}}[m, n] e^{j \frac{2\pi m l}{M}}. \label{xtd2}
    \vspace{-2mm}}
    \end{equation}
    
    \begin{figure}[t]
    \centering
    \vspace{-3mm}
    \includegraphics[width=1\linewidth]{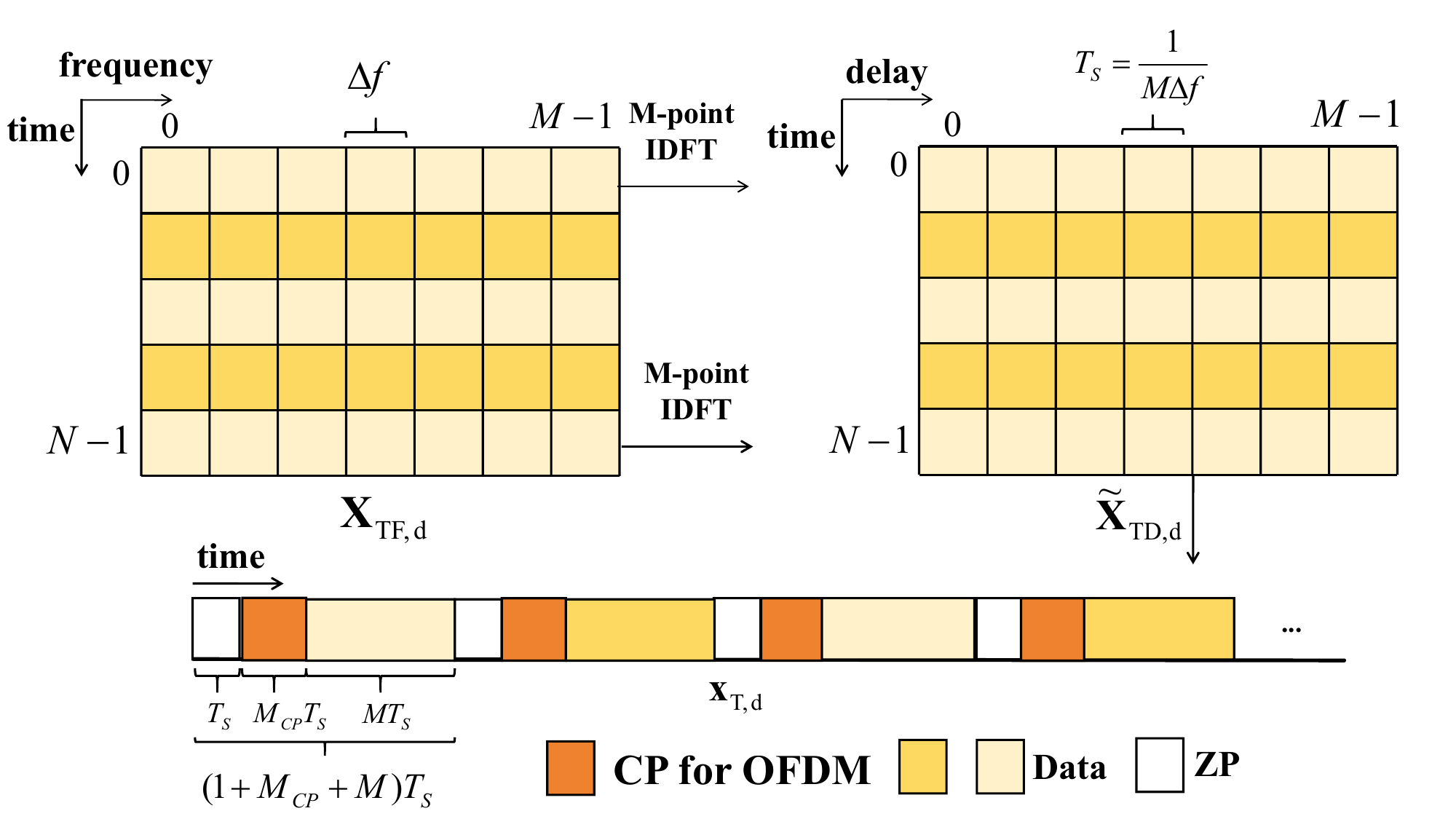}\vspace{-2mm}  
    \caption{Data symbols in TF-, TD- and time-domains.}
    \label{TF}
    \vspace{-3mm}
    \end{figure}
    \begin{figure}[t]
    \vspace{-1mm}
        \centering
        \includegraphics[width=1\linewidth]{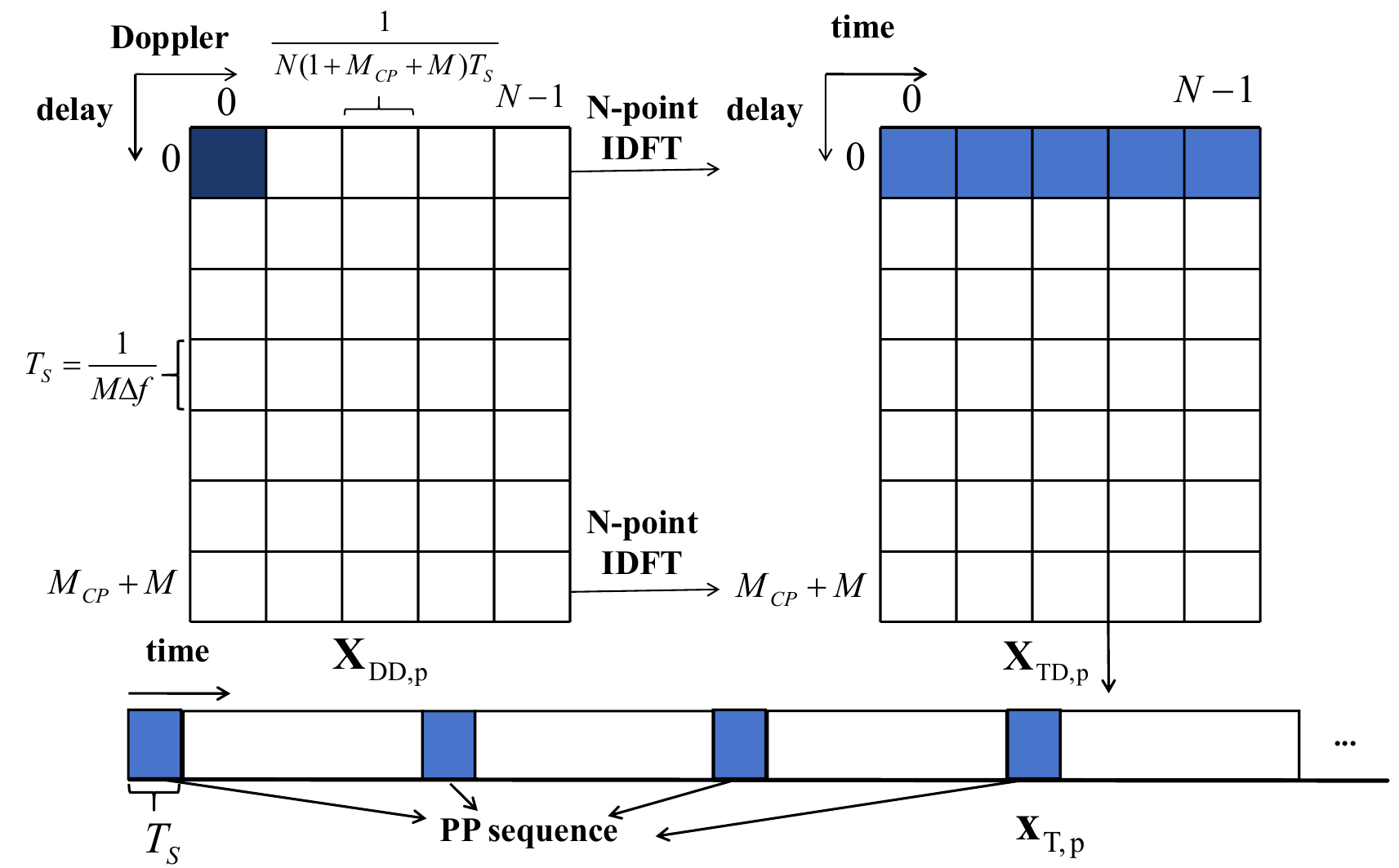}\vspace{-2mm}  
        \caption{Pilot symbols in DD-, TD-, and time-domains.}
        \label{PP}
        \vspace{-5mm}
    \end{figure}
    
   As shown in Fig. \ref{TF}, similar to conventional OFDM transmission, a CP of length $M_{\mathrm{CP}}$ is appended to each OFDM symbol. To accommodate the inserted time-domain PP sequence converted from the DD-domain, we additionally introduce a ZP segment of length $1$ before the CP. Consequently, the TD-domain signals after CP/ZP insertion are given by
   \vspace{-1mm}\begin{equation}
    \mathbf{X}_{\mathrm{TD,d}}=\mathbf W\mathbf C\tilde{\mathbf{X}}_{\mathrm{TD,d}},
    \label{xTD}
    \vspace{-1mm}\end{equation}
    where the CP insertion matrix
$\mathbf C$ is defined as
\vspace{-2mm}\begin{equation}
  \mathbf{C}=
  \begin{bmatrix}
    \mathbf 0_{M_{\mathrm{CP}}\times (M-M_{\mathrm{CP}})} & \mathbf I_{M_{\mathrm{CP}}} \\
    \;\;\;\;\;\;\;\;\;\;\;\mathbf I_M
  \end{bmatrix},
  \label{C}
\vspace{-2mm}\end{equation}
and the zero-corresponding padding
matrix $\mathbf W$ is defined as
\vspace{-2mm}\begin{equation}
\mathbf{W}
=
\begin{bmatrix}
\mathbf{0}_{1 \times (M+M_{\mathrm{CP}})} \\
\mathbf{I}_{M+M_{\mathrm{CP}}}
\end{bmatrix},
\label{W}
\vspace{-2mm}\end{equation}
with $M_{\mathrm{De}}=M+M_{\mathrm{CP}}+1$ denoting the number of time samples within each extended OFDM symbol. After P/S conversion, the TD-domain signals are transformed to the time-domain, and we have $\mathbf{x}_{\mathrm{T,d}}[nM_{\mathrm{De}}+l] = 0$, if $l = 0$, otherwise 
\vspace{-2mm}
    \begin{equation}
    \label{xTd}
    \mathbf{x}_{\mathrm{T,d}}[nM_{\mathrm{De}}+l]= \frac{1}{\sqrt{M}} \sum\limits_{m=0}^{M-1}\mathbf{X}_{\mathrm{TF,d}}[m, n] e^{j \frac{2\pi m [ l-M_{\mathrm{CP}}-1 ]_M}{M}},\vspace{-1mm}
    \end{equation} 
with \(l \in \{1, \ldots, M_{\mathrm{De}}-1\}\) and \(n \in \{0, 1, \ldots, N-1\}\).

    The transmitted DD-domain pilot symbols $\mathbf{X}_{\mathrm{DD,p}} \in \mathbb{C}^{M_{\mathrm{De}} \times N}$ are indexed by $l=0,1,...,M_{\mathrm{De}}-1$ and $k=0,1,...,N-1$. As illustrated in Fig.~\ref{PP}, the single complex pilot symbol is placed at the $(0,0)$ position in the DD-domain, while all the remaining elements are set to zero, i.e.,
    \vspace{-2mm}
    \begin{align}
    \mathbf{X}_{\mathrm{DD,p}}[l, k] = \begin{cases} x_\mathrm{p}, & \text{if } l = 0,k=0 \\0, & \text{else}
\end{cases}
\label{xDDp}
    \end{align}\par\vspace{-2mm}\noindent
    where $|x_\mathrm{p}|=\sqrt{{E}_\mathrm{P}}$ and ${E}_\mathrm{P}$ represents the pilot power.

    Then, an $N$-point IDFT is performed along the Doppler dimension to transform the pilot symbol from the DD-domain to the TD-domain:
    \vspace{-2mm}
    \begin{equation}  \mathbf{X}_{\mathrm{TD,p}}[l,n]=\frac{1}{\sqrt{N}} \sum\nolimits_{k=0}^{N-1} \mathbf{X}_{\mathrm{DD,p}}[l,k] e^{j \frac{2\pi k n}{N}},
    \label{xTp1}\vspace{-2mm}
        \end{equation}
    which are further arranged into the time-domain via P/S conversion, i.e., 
    \vspace{-2mm}
    \begin{align}
    \mathbf{x}_{\mathrm{T,p}}[nM_{\mathrm{De}}+l]=\begin{cases} x_\mathrm{pp}, & \text{if } l = 0\\0, & \text{else}
    \end{cases}
    \label{x_T_p}
    \end{align}\par\vspace{-2mm}\noindent
    with $x_\mathrm{pp}=\frac{x_\mathrm{p}}{\sqrt{N}}$. As can be seen from \eqref{x_T_p} and Fig. \ref{PP}, a single pulse in the DD-domain manifests as a periodic pulse train in the time-domain, which we refer to as pulsone pilot (PP) sequence in this paper. Let $E_\mathrm{pp}=|x_\mathrm{pp}|^2$ represents the pulsone pilot power, then $E_\mathrm{pp}=\frac{E_\mathrm{p}}{N}$ holds. To characterize the impact of the pilot signal on the PAPR of the time domain transmitted waveform, the PP-to-data ratio (PDR) is considered in this paper, which is defined as the ratio between the average time-domain power of the PP and that of the OFDM symbols, i.e., $\mathrm{PDR} = \frac{E_\mathrm{pp}}{E_\mathrm{s}}$, with $E_\mathrm{s}$ denotes the power of the TF-domain data symbols.
    
\begin{figure}[t]
        \centering\vspace{-1mm}
        \includegraphics[width=1\linewidth]{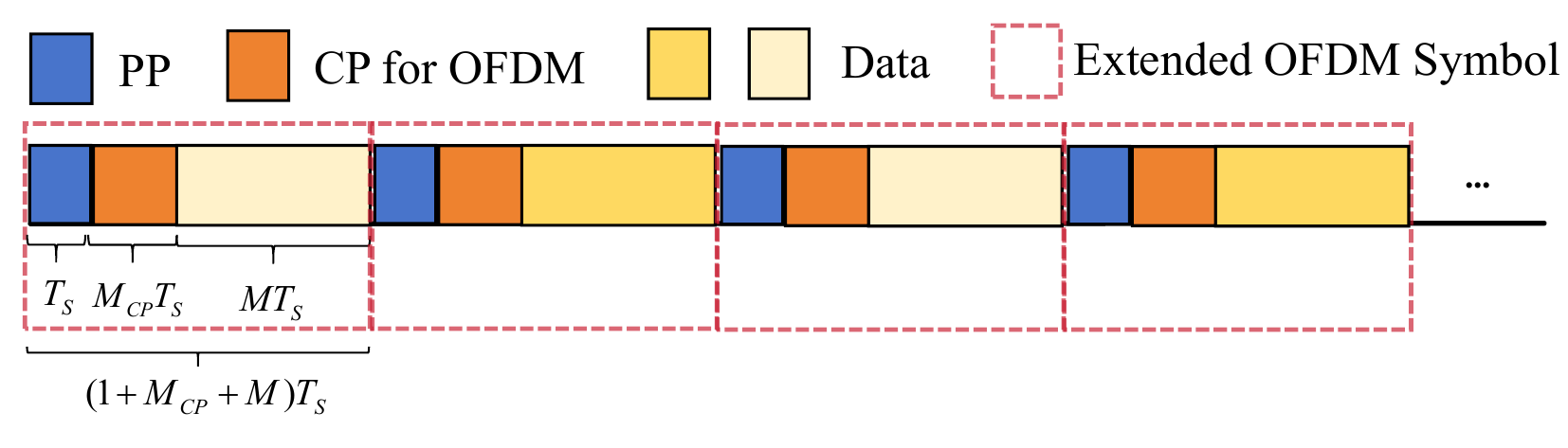}\vspace{-2mm}  
        \caption{The transmitted time-domain signals}
        \label{TX}
        \vspace{-5mm}
    \end{figure}

    Then, by superimposing the time-domain PP sequence onto the data signal sequence, the overall time-domain discrete sample is given by
    \vspace{-1mm}\begin{equation}
    \mathbf{x}_{\mathrm{T}}[nM_{\mathrm{De}}+l] = \mathbf{x}_{\mathrm{T,p}}[nM_{\mathrm{De}}+l]+\mathbf{x}_{\mathrm{T,d}}[nM_{\mathrm{De}}+l].\label{eq:st}\vspace{-1mm}
    \end{equation}
    As illustrated in Fig. \ref{TX}, each extended OFDM symbol therefore spans a duration of $T=M_{\mathrm{De}} T_s$. We note that the proposed DD-a-OFDM transceiver architecture not only enables the PP sequence to be compatible with the existing OFDM waveform structure, but also avoids introducing additional interference to the TF-domain data symbols, as will be demonstrated later in section \ref{pattern}. Then, they are pulse-shaped using a transmit filter \(g_{\mathrm{tx}}(t)\) to form the continuous-time transmit signal
    \vspace{-2mm}
    \begin{equation}
    s(t) = \sum_{l=0}^{M_{\mathrm{De}}-1}\sum_{n=0}^{N-1}\mathbf{x}_{\mathrm{T}}[nM_{\mathrm{De}}+l]\, g_{\mathrm{tx}}(t - lT_s - nT).\vspace{-2mm}
    \label{st2}
    \end{equation}
    
    \vspace{-2mm}
    \subsection{\textcolor{black}{The Received Signal Model}}
    After going through a doubly-selective fading channel, the received signal $r(t)$ is given by
    \vspace{-2mm}
    \begin{equation}
    r(t) = \int \int h(\tau, \nu) s(t - \tau) e^{j 2 \pi \nu (t - \tau)} \, d\tau \, d\nu + w(t),
    \label{rt}\vspace{-2mm}
    \end{equation}
    where $w(t)$ denotes the additive white Gaussian noise with zero mean and power spectral density $N_0$. In \eqref{rt}, the DD-domain channel response is given by 
    \vspace{-2mm}
    \begin{equation}
    h(\tau, \nu) = \sum\nolimits_{i=1}^{P} g_i \delta(\tau - \tau_i) \delta(\nu - \nu_i), \vspace{-2mm}
    \end{equation}
    where $\delta(\cdot)$ is the Dirac delta function, $P$ is the number of propagation
    paths, $g_i$, $\tau_i$ and $\nu_i$ denote the complex path gain, delay and
    Doppler shift associated with the $i$-th path, respectively. \textcolor{black}{Since the time-domain sampling interval $\frac{1}{M\Delta f}$ in wideband systems is typically small, the impact of fractional delay can be neglected \cite{tsefundamental}. Therefore, an integer-delay channel model is adopted, while the fractional Doppler shift is taken into account \cite{cross_domain}, i.e.,
    $\tau_i = \frac{l_i}{M \Delta f}$ and $\nu_i = \frac{k_i+\kappa_i}{N' T}$, where $\kappa_i \in (-\tfrac{1}{2},\tfrac{1}{2}]$ denotes the fractional Doppler component, and $N'<N$ denotes the number of consecutive extended OFDM symbols received for CE and data detection\footnote{Note that in the proposed DD-a-OFDM scheme, it is not necessary to receive all the transmitted $N$ symbols for CE and data detection. In fact, receiving signal processing is performed based on the first $N'$ received extended OFDM symbols. A larger $N'$ implies a higher Doppler resolution and a higher system latency.}.
    Let $l_{\max}$ and $k_{\max}$ denote the maximum delay and Doppler tap, respectively. Furthermore, the channel is assumed to be underspread, i.e., $l_{\max} \ll M$ and $k_{\max} \ll N'$.}
    
    Then, $r(t)$ is passed through a matched filter \(g_{\mathrm{rx}}(t)\) and samples at \( t = n'T + l'T_s \), with \( n' \in \{0, 1, \dots, N'-1\} \) and \( l' \in \{0, \dots, M_{\mathrm{De}}-1\} \), yielding:
    \begin{align}
    &\mathbf{y}_{\mathrm{T}}[n'M_{\mathrm{De}}+l'] \hspace{-1mm}
    = \hspace{-1mm}\sum\nolimits_{i=1}^{P} g_i  \hspace{-1mm}
    \sum\nolimits_{l=0}^{M_{\mathrm{De}}-1}\hspace{-1mm}\sum\nolimits_{n=0}^{N'-1}\mathbf{x}_{\mathrm{T}}[nM_{\mathrm{De}}+l]\, \notag \\
    &\cdot\; e^{j 2\pi \nu_i (nT + lT_s)} A_{g_{\mathrm{tx}}, g_{\mathrm{rx}}} \Big( 
    (n' - n)T + (l'-l)T_s - \tau_i, \, \nu_i 
    \Big)\notag\\
    &+\mathbf{w}_{\mathrm{T}}[n'M_{\mathrm{De}}+l'],
    \label{eq:yTD_full}
    \end{align}
    \par\vspace{-1mm}\noindent
    where $\mathbf{w}_{\mathrm{T}} \sim \mathcal{CN}(\mathbf{0}, N_0 \mathbf{I}_{N'M_{\mathrm{De}}})$ denotes the discrete noise sample after matched filtering. And \( A_{g_{\mathrm{tx}}, g_{\mathrm{rx}}}(\tau, \nu) \) is the cross-ambiguity function between the transmit and receive filters
    \vspace{-2mm}
    \begin{equation}
    A_{g_{\mathrm{tx}}, g_{\mathrm{rx}}}(\tau, \nu) = \int g_{\mathrm{tx}}(u) \, g_{\mathrm{rx}}(\tau - u) \, e^{j2\pi \nu u} \, du, \label{eq:ambiguity}\vspace{-2mm}
    \end{equation} 
    which can be approximated as
    \vspace{-2mm}\begin{equation}
    A_{g_{\mathrm{tx}}, g_{\mathrm{rx}}}(\tau, \nu) \approx e^{j2\pi \nu \tau} \int g_{\mathrm{tx}}(u) g_{\mathrm{rx}}(\tau - u) \, du, \label{eq:amb_approx}\vspace{-2mm}
    \end{equation}
    when \( T_s \) is sufficiently small. If \( g_{\mathrm{tx}}(t) \) and \( g_{\mathrm{rx}}(t) \) are designed such that their convolution is concentrated within a narrow support (e.g., using root-raised-cosine pulses), the integral becomes negligible for \(|\tau| > T_s/2\), and thus the ambiguity function in \eqref{eq:yTD_full} becomes sharply peaked at \( n = n' \) and \( l = l' - l_i \).  The received time-domain sampled signal is given by\footnote{When $\tau_i$ is off-grid with respect to the delay resolution, 
    a closed-form expression of $A_{g_{\mathrm{tx}}, g_{\mathrm{rx}}}
    \!\left(t - nT - lT_s - \tau_i, \nu_i\right)$ 
    can still be derived by oversampling the transceiver pulses \cite{MP,off_grid}. 
    However, for the sake of simplicity, this case is not considered in the present work.}
    \vspace{-1mm}
    \begin{align}
    \mathbf{y}_{\mathrm{T}}[n'M_{\mathrm{De}}+l'] 
    =\sum\nolimits_{i=1}^{P} g_i e^{j 2\pi \frac{(k_i+\kappa_i) (n'M_{\mathrm{De}} + l'-l_i)}{N'M_{\mathrm{De}}}} \notag\\
    \cdot \mathbf{x}_{\mathrm{T}}[n'M_{\mathrm{De}}+l'-l_i]+\mathbf{w}_{\mathrm{T}}[n'M_{\mathrm{De}}+l'].
    \label{yT}
    \end{align}

In the proposed transceiver architecture, CE relies on the DD-domain pilot. Thus, in the DD-domain branch, only $l_{\rm max}+1$ symbols starting from each PP symbol, i.e., $\mathbf{y}_{\mathrm{T}}[n'M_{\mathrm{De}}+l'],n'=0,1,..,N'-1,l'=0,1,...,l_{\rm max}$, are transformed back to the DD-domain through an $N'$-point DFT, yielding
\vspace{-2mm}
\textcolor{black}{
\begin{equation}
\mathbf{Y}_{\mathrm{DD}}[l', k'] = \frac{1}{\sqrt{N'}} \sum\nolimits_{n'=0}^{N'-1}\mathbf{y}_{\mathrm{T}}[n'M_{\mathrm{De}}+l'] \, e^{-j \frac{2\pi n' k'}{N}},
\label{yDD}
\end{equation}}
\par\vspace{-2mm}\noindent
with \(k' \in \{0, \ldots, N'-1\}\), $l' \in \{0, \ldots, l_{\rm max}\}$.

Furthermore, the time-domain signals corresponding to the transmitted data symbols are converted back to the TF-domain for symbol detection. In the TF-domain branch, after discarding the CP and PP, applying an $M$-point DFT yields the received TF-domain symbols as
\vspace{-2mm}
\textcolor{black}{
\begin{align}
\mathbf{Y}_{\mathrm{TF}}[m', n'] = &\frac{1}{\sqrt{M}} \sum\nolimits_{l'=0}^{M-1}\mathbf{y}_{\mathrm{T}}[n'M_{\mathrm{De}}+l'+M_{\mathrm{CP}}+1] \notag\\
& \cdot\, e^{-j \frac{2\pi m' l'}{M}},\label{yTF}
\end{align}}
\hspace*{-0.7em} with \(n' \in \{0, 1, \ldots, N'-1\}\), $m' \in \{0, 1, \ldots, M-1\}$.

    \begin{remark}
    Besides its compatibility with OFDM, the proposed DD-a-OFDM transceiver architecture enjoys the following three benefits:\\
    1) \textit{\textbf{Low PAPR}}: Although the pilot power is highly concentrated in the DD-domain, its time-domain representation, i.e., PP sequence, spreads this power across $N$ OFDM symbols. Consequently, each PP symbol contains only $\frac{1}{N}$ of the original DD-domain pilot power. This structural property preserves the high pilot power in the DD-domain for accurate CE while avoiding a potential increase in the PAPR of the transmitted time-domain OFDM waveform. \\
    2) \textit{\textbf{High spectral efficiency}}: To combat both frequency-selective and time-selective fading, conventional OFDM systems typically insert pilots across subcarriers and OFDM symbols in the TF-domain \cite{COFDM2}. Let $\alpha$ denote the pilot overhead ratio, the effective spectral efficiency is given by $\frac{M(1-\alpha)}{M_{\mathrm{De}}-1} \log |\mathcal{A}| \; \text{bits/s/Hz}$. \textcolor{black}{In contrast, the proposed PP scheme avoids the use of conventional OFDM pilot symbols and requires only one pilot per extended OFDM symbol}, i.e., the effective spectral efficiency is given by $\frac{M}{M_{\mathrm{De}}} \log |\mathcal{A}| \; \text{bits/s/Hz}$. Notably, in conventional OFDM systems, $\alpha$ typically increases as the channel becomes more severely doubly-selective fading, thereby increasing pilot overhead and decreasing spectral efficiency. In contrast, the proposed PP scheme significantly reduces pilot overhead, thereby improving spectral efficiency. \textcolor{black}{Furthermore, since the proposed DD-a-OFDM scheme adopts a DD-domain pilot design embedded within the conventional OFDM signal structure rather than a fully DD-domain transceiver scheme, it does not require additional DD-domain guard intervals to avoid pilot-to-data and data-to-pilot interference, as in conventional OTFS schemes\cite{yuan_data_aided,off_grid,Thre2,Thre1}. For large delay spreads, a longer CP is required to preserve the circular convolution property of OFDM, resulting in additional CP overhead and spectral efficiency degradation for both the OFDM and DD-a-OFDM. Such degradation is an inherent limitation of OFDM systems rather than being additionally introduced by the proposed DD-a-OFDM pilot design. On the other hand, since no additional Doppler-domain guard interval is required, the spectral efficiency does not decrease with increasing Doppler spread.}\\
    3) \textit{\textbf{Flexible latency control and sequential reception}}: As mentioned before, the proposed DD-a-OFDM architecture can adaptively choose $N'<N$, depending on the Doppler resolution requirement, which allows for flexible latency control. Moreover, the proposed scheme naturally enables sequential receiving signal processing. Specifically, initial channel estimates and detected symbols are obtained from limited observations (i.e., the first $N'$ symbols) and then used as prior information for subsequent CE and detection as more signals are received, thereby progressively improving CE accuracy and detection performance. Since this work mainly focuses on the proposed DD-a-OFDM transceiver architecture and iterative algorithm design, this aspect is not further explored.
    \end{remark}

\vspace{-3mm}
	\section{\textcolor{black}{Input-Output Relationship in the TF- and DD-domains}}
    \label{pattern}

    In this section, for the proposed DD-a-OFDM transceiver architecture, we analyze the input-output relationship in both the DD- and TF-domains, laying the foundation for the subsequent cross-domain iterative algorithm design.

\vspace{-2mm}
\subsection{\textcolor{black}{TF-domain Input-Output Relationship}}
By substituting \eqref{eq:st} and \eqref{yT} into \eqref{yTF}, the TF-domain received signal is given by \eqref{YTFmn} at the top of the next page.
    \begin{figure*}[!t] 
        \centering
        \vspace*{-4mm}
\begin{align}
\label{YTFmn}
\mathbf{Y}_{\mathrm{TF}}[m', n'] &=\underbrace{\frac{1}{\sqrt{M}}\sum\nolimits_{i=1}^{P} g_i \sum\nolimits_{l'=0}^{M-1} 
\mathbf{x}_{\mathrm{T,d}}[n'M_{\mathrm{De}}+l'+M_{\mathrm{CP}}+1-l_i]\, e^{j 2\pi \frac{(k_i+\kappa_i) (n'M_{\mathrm{De}} + l'+M_{\mathrm{CP}}+1-l_i)}{N'M_{\mathrm{De}}}}e^{-j \frac{2\pi m' l'}{M}}}_{\substack{\mathbf{Y}_{\mathrm{TF,d}}[m', n']:\;\;\text{Desired\, signal\, in\, the \, TF\, domian}}}\\
&\hspace{-5mm}+\underbrace{\frac{1}{\sqrt{M}}\sum\nolimits_{i=1}^{P} g_i \sum\nolimits_{l'=0}^{M-1} 
\mathbf{x}_{\mathrm{T,p}}[n'M_{\mathrm{De}}+l'+M_{\mathrm{CP}}+1-l_i]\, e^{j 2\pi \frac{(k_i+\kappa_i) (n'M_{\mathrm{De}} + l'+M_{\mathrm{CP}}+1-l_i)}{N'M_{\mathrm{De}}}}e^{-j \frac{2\pi m' l'}{M}}}_{\mathbf{I}_{\mathrm{TF},p}[m', n']:\;\;\substack{\text{Interference\, in\,the\, TF\, domian}}}+\mathbf{W}_{\mathrm{TF}}[m',n']. \notag
\end{align}
\par\vspace{-2mm}\noindent
    \hrulefill 
    \end{figure*}
\textcolor{black}{Here, $\mathbf{I}_{\mathrm{TF,p}}$ represents the interference signal in the TF-domain caused by the DD-domain pilot, $\mathbf{Y}_{\mathrm{TF,d}}$ consists of the desired TF-domain data symbols only, and $\mathbf{W}_{\mathrm{TF}}[m', n']=\frac{1}{\sqrt{M}} \sum_{l'=0}^{M-1}\mathbf{w}_{\mathrm{T}}[n'M_{\mathrm{De}}+l'+M_{\mathrm{CP}}+1] \, e^{-j \frac{2\pi m' l'}{M}}$ denotes the effective TF-domain noise.}


In fact, owing to the CP guard interval between PP and data symbols in each extended OFDM symbol, there is no interference from pilot to data. Specifically, since $M_{\mathrm{CP}} > l_{\max}$, we have $l' + M_{\mathrm{CP}} + 1 - l_i \ge 1$ always holds for $l'=0,1,...,M-1$. Recalling \eqref{x_T_p}, we further observe that $\mathbf{x}_{\mathrm{T,p}}[n'M_{\mathrm{De}}+l'+M_{\mathrm{CP}}+1-l_i]=0$ always holds for \(n' \in \{0, 1, \ldots, N'-1\}\) and \(l' \in \{0, 1, \ldots, M-1\}\), i.e., $\mathbf{I}_{\mathrm{TF},p}[m', n']=0$, which implies that there is no interference from the PP pilot to data symbols in the TF-domain.


Besides, owing to the presence of CP, 
$\mathbf{x}_{\mathrm{T,d}}[n'M_{\mathrm{De}}+l'+M_{\mathrm{CP}}+1-l_i]=\tilde{\mathbf{X}}_{\mathrm{TD,d}}\left[[l'-l_i]_M,n'\right]$ always holds. Thus, $\mathbf{Y}_{\mathrm{TF}}$ can be rewritten as
\vspace{-2mm}\textcolor{black}{
\begin{align}
\label{YTF1}
\hspace{-1mm}\mathbf{Y}_{\mathrm{TF}}&[m', n'] \hspace{-1mm}
=\hspace{-1mm}\frac{1}{\sqrt{M}}\hspace{-1mm}\sum_{i=1}^{P} g_i \hspace{-1mm}\sum_{l'=0}^{M-1} 
\tilde{\mathbf{X}}_{\mathrm{TD,d}}\left[[l'-l_i]_M,n'\right]\, e^{-j \frac{2\pi l'b}{M}} \notag \\
&\cdot e^{j 2\pi \frac{(k_i+\kappa_i) (n'M_{\mathrm{De}} + l'+M_{\mathrm{CP}}+1-l_i)}{N'M_{\mathrm{De}}}}+\mathbf{W}_{\mathrm{TF}}[m',n'].
\end{align}}
\par\vspace{-2mm}\noindent
Combining \eqref{xtd2} and \eqref{YTF1}, let $\mathbf{Y}_{\mathrm{TF}}=[\mathbf{y}^{(0)}_{\mathrm{TF}},\mathbf{y}^{(1)}_{\mathrm{TF}} ,...,\mathbf{y}^{(N'-1)}_{\mathrm{TF}}  ] $, $\mathbf{X}_{\mathrm{TF,d}}=[\mathbf{x}^{(0)}_{\mathrm{TF,d}},\mathbf{x}^{(1)}_{\mathrm{TF,d}} ,...,\mathbf{x}^{(N'-1)}_{\mathrm{TF,d}}  ]$ and $\mathbf{W}_{\mathrm{TF}}=[\mathbf{w}^{(0)}_{\mathrm{TF}},\mathbf{w}^{(1)}_{\mathrm{TF}} ,...,\mathbf{w}^{(N'-1)}_{\mathrm{TF}}  ]$, the TF-domain input-output relationship can be further expressed in matrix form as
\begin{equation}
\mathbf{y}^{(n')}_{\mathrm{TF}} =\mathbf{H}_{\text{TF}}^{(n')} \mathbf{x}^{(n')}_{\mathrm{TF,d}} + \mathbf{w}_{\mathrm{TF}}^{(n')}, \label{eq:matrix_form}\vspace{-1mm}
\end{equation}
where the effective TF-domain channel for the $n'$-th OFDM symbol is given by
\begin{equation}
\label{H_TF_d}
\mathbf{H}_{\text{TF}}^{(n')}\hspace{-1mm}
=\hspace{-1mm} \sum_{i=1}^P g_ie^{j 2\pi(k_i+\kappa_i)\left(\frac{n'}{N'}+\frac{M_{\mathrm{CP}}+1-l_i}{N'M_{\mathrm{De}}}\right) }\mathbf{F}_M\mathbf{\Delta}_M^{k_i + \kappa_i}\mathbf{\Pi}_M^{l_i}\mathbf{F}_M^{\text{H}},
\end{equation}
where the permutation matrix $\mathbf{\Pi}_M=\operatorname{circ}([0,1,0,\dots,0]^{\mathrm{T}})\in\mathbb{R}^{MN'\times MN'}$ models the delay effect, the diagonal matrix $\mathbf{\Delta}_M=\operatorname{diag}(z^0,z^1,\dots,z^{M-1})\in\mathbb{C}^{M\times M}$ captures the Doppler effect within each OFDM symbol, with $z=e^{\frac{j2\pi}{N'M_{\mathrm{De}}}}$, and $e^{j2\pi(k_i+\kappa_i)\left(\frac{n'}{N'}+\frac{M_{\mathrm{CP}}+1-l_i}{N'M_{\mathrm{De}}}\right)}$ in \eqref{H_TF_d} represents the initial phase experienced by the first data symbol of the $n'$-th OFDM symbol.
Note that in \eqref{eq:matrix_form}, the effective TF-domain channel $\mathbf{H}_{\text{TF}}^{(n')}$ is not diagonal in high-mobility channels, where the ICI channels are characterized by the off-diagonal entries in $\mathbf{H}_{\text{TF}}^{(n')}$.

\subsection{\textcolor{black}{DD-domain Input-Output Relationship}}
\label{DDI}
\textcolor{black}{Since CE is performed in the DD domain, the received signal associated with the pilot symbol is the desired signal, while the contributions from the TF-domain data symbols act as interference. By substituting \eqref{eq:st} and \eqref{yT} into \eqref{yDD}, the DD-domain received signal is given by \eqref{Y_DD} at the top of this page,} 
    \begin{figure*}[!t] 
        \centering
        \vspace*{-7mm}

\begin{align}
\label{Y_DD}
\mathbf{Y}_{\mathrm{DD}}[l', k'] 
&=\underbrace{\frac{1}{\sqrt{N'}}\sum\nolimits_{i=1}^{P} g_i \sum\nolimits_{n'=0}^{N'-1}\,
\mathbf{x}_{\mathrm{T,d}}[n'M_{\mathrm{De}}+l'-l_i]\,  e^{j 2\pi \frac{(k_i+\kappa_i) (n'M_{\mathrm{De}} + l'-l_i)}{N'M_{\mathrm{De}}}}e^{-j \frac{2\pi n' k'}{N'}}}_{\mathbf{I}_{\mathrm{DD,d}}[l',k']:\;\substack{\text{Interference\, in\, the\, DD\, domian}}}\notag\\
&\hspace{-5mm}+\underbrace{\frac{1}{\sqrt{N'}}\sum\nolimits_{i=1}^{P} g_i \sum\nolimits_{n'=0}^{N'-1}\,
\mathbf{x}_{\mathrm{T,p}}[n'M_{\mathrm{De}}+l'-l_i]\, e^{j 2\pi \frac{(k_i+\kappa_i) (n'M_{\mathrm{De}} + l'-l_i )}{N'M_{\mathrm{De}}}}e^{-j \frac{2\pi n' k'}{N'}}}_{\mathbf{Y}_{\mathrm{DD,p}}[l',k']:\;\substack{\text{Desired\, signal\, in\, the\, DD\, domian}}}+\mathbf{W}_{\mathrm{DD}}[l', k'],
\end{align}
\par\vspace{-2mm}\noindent
    \hrulefill 
    \vspace{-2mm}
    \end{figure*}
\textcolor{black}{where $\mathbf{Y}_{\mathrm{DD,p}}$, $\mathbf{I}_{\mathrm{DD,d}}$, and $\mathbf{W}_{\mathrm{DD}}=\operatorname{vec}^{-1}(\mathbf{w}_{\mathrm{T}})\mathbf{F}_{N'}$ denote the desired signal, data-induced interference, and effective DD-domain noise, respectively.}

By substituting \eqref{x_T_p}, the desired DD-domain signal $\mathbf{Y}_{\mathrm{DD,p}}$ is given by \cite{pulseshape,MP}
\vspace{-2mm}
\begin{align}
&\mathbf{Y}_{\mathrm{DD,p}}[l',k']     \label{DDrela}\\
&=
\frac{1}{\sqrt{N'}}\sum_{i=1}^{P}
g_i\,x_{\mathrm{pp}}\,
    \sum_{{n'}=0}^{N'-1}e^{-j2\pi n'\frac{k'-(k_i+\kappa_i)}{N'} } \delta([l'-l_i]_{M_{\mathrm{De}}}).\notag
    \vspace{-2mm}
\end{align}
Specifically, when the Doppler shifts are integer-valued, i.e., $\kappa_i=0$, \eqref{DDrela} can be further expressed as
\vspace{-2mm}
\begin{equation}
\mathbf{Y}_{\mathrm{DD,p}}[l',k']
=\sqrt{N'}\sum_{i=1}^{P}
g_i\,x_{\mathrm{pp}}\,
    \delta([l'-l_i]_{M_{\mathrm{De}}})\delta([k'-k_i]_{N'}),
    \label{DDrela_intger}
    \vspace{-1mm}
\end{equation}
\textcolor{black}{which reduces to the conventional OTFS twisted 2D circular convolution input–output relationship \cite{pulseshape}. It is worth noting that, in \eqref{DDrela_intger}, when $l'=l_i$ and $k'=k_i$, the corresponding received symbol power is $N'\sigma_i^2 E_{\mathrm{pp}}$, where $\mathbb{E}{|g_i|^2}=\sigma_i^2$ represents the power of the $i$-th propagation path. This demonstrates that the $N'$ received time-domain PP signals corresponding to the $i$-th path are coherently combined in the DD-domain. Even in fractional Doppler channel with Doppler-domain energy leakage in \eqref{DDrela}, most of the energy remains concentrated around the corresponding DD bin, thereby preserving the coherent combining gain to a large extent.}

\label{gaussian}
Next, we analyze the interference signal in the DD-domain caused by the TF-domain data symbols. Substituting \eqref{xTd} into \eqref{Y_DD}, the DD-domain interference signal $\mathbf{I}_{\mathrm{DD,d}}$ is given by
\vspace{-1mm}
\begin{equation}
\label{Y_DD_I}
\mathbf{I}_{\mathrm{DD,d}}[l',k'] 
=\sum\nolimits_{i=1}^{P} g_i \,f_{l',k',i}, \vspace{-1mm}
\end{equation}
   \begin{figure*}[!t] 
        \centering
        \vspace{-5mm}
    \begin{align}\textcolor{black}{
    \label{flki}
    f_{l',k',i}=\begin{cases} 0, & \text{if } l' = l_i\\ \frac{1}{\sqrt{MN}} \sum\nolimits_{m'=0}^{M-1}\sum\nolimits_{n'=1}^{N'-1}\,e^{j 2\pi \frac{(k_i+\kappa_i) (n'M_{\mathrm{De}}+l'-l_i)}{M_{\mathrm{De}}N'}}e^{-j 2\pi\frac{ n' k'}{N'}} e^{j2\pi  \frac{m' [M+l'-l_i]_M}{M}}\mathbf{X}_{\mathrm{TF,d}}[m',n'-1], & \text{if } l' < l_i\\
    \frac{1}{\sqrt{MN}} \sum\nolimits_{m'=0}^{M-1}\sum\nolimits_{n'=0}^{N'-1}\,e^{j 2\pi \frac{(k_i+\kappa_i) (n'M_{\mathrm{De}}+l'-l_i)}{M_{\mathrm{De}}N'}}e^{-j 2\pi\frac{ n' k'}{N'}} e^{j2\pi  \frac{m' [l'-l_i-M_{\mathrm{CP}}-1]_M}{M}}\mathbf{X}_{\mathrm{TF,d}}[m',n'], & \text{if } l' > l_i
    \end{cases}}
    \end{align} 
\par\vspace{-2mm}\noindent
     \hrulefill 
     \vspace{-5mm}
    \end{figure*}
where $f_{l',k',i}$ denotes the interference signal at the $[l',k']$-th DD bin caused by the TF-domain data symbols propagating through the $i$-th path, and is given by \eqref{flki} at the top of next
page.
 
Assuming  $\mathbb{E}[\mathbf{X}_{\mathrm{TF,d}}[m',n']]=0$ for $m'=0,1,\ldots,M-1$ and $n'=0,1,\ldots,N'-1$, $\mathbb{E}[\text{vec}(\mathbf{X}_{\mathrm{TF,d}})(\text{vec}(\mathbf{X}_{\mathrm{TF,d}})))^\mathrm{H}]=E_\mathrm{s}\mathbf{I}_{MN'}$, $\mathbb{E}\{g_i\}=0$ for $i=1,2,\ldots,P$ and $\mathbb{E}\{g_i g_j^*\}=0$ for $i\neq j$. The interference distribution at each DD bin follows $\mathbb{E}\left[\mathbf{I}_{\mathrm{DD,d}}[l',k']\right ]=0$, $\mathbb{E}\left[|\mathbf{I}_{\mathrm{DD,d}}[l',k']|^2\right ]=\sum_{i=1}^{P} \sigma_i^2\;\mathbb{E}\left[|f_{l',k',i}|^2\right]$,
    with
   \begin{align}
    \mathbb{E}\left[|f_{l',k',i}|^2\right] =\begin{cases} 0, & \text{if } l' = l_i,\\ \frac{N'-1}{N'}E_\mathrm{s}, & \text{if } l' < l_i,\\
    E_\mathrm{s}, & \text{if } l' > l_i.
    \end{cases}
    \label{fmi2}
    \end{align} 
Each $\mathbf{I}_{\mathrm{DD,d}}[l',k']$ contains the superposition of the received signals corresponding to a large number of TF-domain data symbols. Owing to this aggregation effect, by the central limit theorem, the resulting interference signal follows a circularly symmetric complex Gaussian distribution, i.e., $\mathbf{I}_{\mathrm{DD,d}}[l',k']
\sim
\mathcal{CN}\!\left(0,
\sum_{i=1}^{P} \sigma_i^2\,\mathbb{E}\!\left[ |f_{l',k',i}|^2 \right]\right).$ \textcolor{black}{In contrast to the coherent combining of the received time-domain PP sequence, the received data symbols exhibit energy spreading and behave approximately as Gaussian noise, which enhances the pilot-to-interference power ratio in the DD-domain observation and facilitates high-accuracy CE.}

    \vspace{-3mm}
	\section{Cross-Domain Iterative Receiver Design}

    In this section, \textcolor{black}{a low-complexity cross-domain iterative reception algorithm is designed for the proposed DD-a-OFDM tranceiver architecture, as shown in Fig. \ref{Iteration}. Without loss of generality, the iteration index is denoted by $(t)$. Specifically, at the $(t)$-th iteration, given the TF-domain detection output $\hat{\mathbf{X}}^{(t-1)}_{\mathrm{TF,d}}$ and channel estimates $\hat{\mathbf{H}}^{(t-1)}_{\mathrm{DD}}$ from the previous iteration, interference reconstruction and cancellation are conducted in the DD-domain. The resulting DD-domain signals after IC $\tilde{\mathbf{Y}}^{(t)}_{\mathrm{DD,d}}$ are then exploited for CE with previous CE error variance matrix ${\mathbf{C}}^{(t-1)}_{\mathrm{DD}}$ and symbol detection error variance matrix ${\mathbf{C}}^{(t-1)}_{\mathrm{TF,d}}$. The obtained channel estimates $\hat{\mathbf{H}}^{(t)}_{\mathrm{DD}}$ and ${\mathbf{C}}^{(t)}_{\mathrm{DD}}$ are in turn fed back to the TF-domain for updating $\hat{\mathbf{X}}^{(t)}_{\mathrm{TF,d}}$ and ${\mathbf{C}}^{(t)}_{\mathrm{TF,d}}$. Note that this cross-domain iterative reception algorithm starts from the DD-domain CE without IC since no prior information on the transmitted TF-domain data symbol is available, i.e., $\hat{\mathbf{X}}^{(0)}_{\mathrm{TF,d}}=\mathbf{0}_{MN'}$. By iteratively exchanging information between the two domains, the DD-domain CE and TF-domain detection performance are progressively improved}.
    
    \textcolor{black}{To facilitate the presentation of the proposed cross-domain iterative reception algorithm, an integer Doppler channel is first considered, i.e., $\kappa_i=0,i=1,2,...,P$, and the extension to the fractional Doppler case will be presented latter. In this case, the DD-domain input–output relationship in \eqref{DDrela_intger} can be simplified as}
    \vspace{-2mm}
    \begin{equation}
    \mathbf{Y}_{\mathrm{DD}}[l',k'] 
    =
    h_{l',k'}\sqrt{N'}x_{\mathrm{pp}}+\mathbf{I}_{\mathrm{DD,d}}[l',k'] +\mathbf{W}_{\mathrm{DD}}[l',k'].
    \label{eq:DD_IO_effective}
    \end{equation}
    where $\mathbf{H}_{\mathrm{DD}}=[\mathbf{h}_0,\mathbf{h}_1,...,\mathbf{h}_{l_{\rm max}}]^\mathrm{T}\in\mathbb C^{(l_{\rm max}+1)\times (2k_{\rm max}+1)}$, 
    $\mathbf{h}_{l'}=[h_{l',0},...,h_{l',k_{\rm max},},h_{l',N'-k_{\rm max}},...,h_{l',N'-1}]^\mathrm{T}$ with $h_{l',k'}=\sum_{i=1}^{P} g_i\delta([l' - l_i]_{M_{\mathrm{De}}})\delta([k' - k_i]_{N'})$,
    and
    \vspace{-2mm}
    \begin{equation}
    \mathbf{I}_{\mathrm{DD,d}}[l',k'] =
    \sum_{l=0}^{l_{\rm max}}\sum_{k\in\mathcal D}h_{l,k}\,f_{l',k',l,k}\vspace{-2mm}
    \label{YDDd}
    \end{equation}
    with $\mathcal{D}=\{0,1,\ldots,k_{\max}\}\cup\{N'-k_{\max},\ldots,N'-1\}$. Similar to analysis in Section \ref{DDI}, we have $\mathbb{E}\left[\mathbf{I}_{\mathrm{DD,d}}[l',k']\right ]=0$ and 
    \vspace{-2.5mm}\begin{equation}
    \textcolor{black}{\mathbb{E}\left[|\mathbf{I}_{\mathrm{DD,d}}[l',k']|^2\right ]=\sum_{l=0}^{l_{\rm max}}\sum_{k\in\mathcal D}\mathbb{E}|h_{l,k}|^2\;\mathbb{E}\left[|f_{l',k',l,k}|^2\right],}
    \label{IDDdlk}\vspace{-2mm}
    \end{equation}
    where $f_{l',k',l,k}$ is introduced to replace $f_{l',k',i}$ in \eqref{Y_DD_I} in the integer delay and Doppler channel, eliminating the explicit dependence on $P$ and enabling per-DD-bin CE and processing at the receiver, which is defined in \eqref{fmi} at the top of next page, 
    \begin{figure*}[!t] 
        \centering
        \vspace*{-7mm}
    \begin{align}
    \textcolor{black}{
    f_{l',k',l,k}=\begin{cases} 0, & \text{if } l=l',\\
    \frac{1}{\sqrt{MN}} \sum\nolimits_{m'=0}^{M-1}\sum\nolimits_{n'=1}^{N'-1}\,e^{j 2\pi \frac{k(n'M_{\mathrm{De}}+l'-l)}{M_{\mathrm{De}}N'}}e^{-j 2\pi\frac{ n' k'}{N'}} e^{j2\pi  \frac{m' [M+l'-l]_M}{M}}\mathbf{X}_{\mathrm{TF,d}}[m',n'-1], & \text{if } l < l',\\
    \frac{1}{\sqrt{MN}} \sum\nolimits_{m'=0}^{M-1}\sum\nolimits_{n'=0}^{N'-1}\,e^{j 2\pi \frac{ k(n'M_{\mathrm{De}}+l'-l)}{M_{\mathrm{De}}N'}}e^{-j 2\pi\frac{ n' k'}{N'}} e^{j2\pi  \frac{m' [l'-l-M_{\mathrm{CP}}-1]_M}{M}}\mathbf{X}_{\mathrm{TF,d}}[m',n'], & \text{if } l > l',
    \end{cases}
    \label{fmi}}
    \end{align} 
    \par\vspace{-2mm}\noindent
    \hrulefill 
    \vspace{-4mm}
    \end{figure*}
    and
    \vspace{-2mm}
    \begin{align}
    \mathbb{E}\left[|f_{l',k',l,k}|^2\right] =\begin{cases} 0, & \text{if } l' = l,\\ \frac{N'-1}{N'}E_\mathrm{s}, & \text{if } l' < l,\\
    E_\mathrm{s}, & \text{if } l' > l.
    \end{cases}
    \label{fmi2}
    \end{align} 
    \par\vspace{-3mm}\noindent
    
    \vspace{-4mm}
	\subsection{DD-domain Interference Cancellation}
        \label{Section:DDIC}

    \begin{figure}[t]
    \centering\vspace{-2mm}        \includegraphics[width=0.9\linewidth]{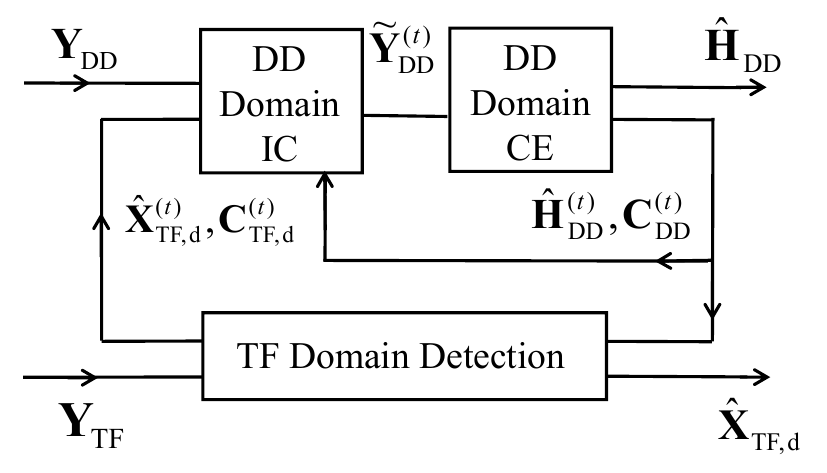}\vspace{-2mm}
    \caption{\textcolor{black}{Cross-domain iterative reception algorithm.}}
    \label{Iteration}\vspace{-4mm}
    \end{figure}

\textcolor{black}{In the $(t)$-th iteration with $t\ge1$, the DD-domain IC is performed using $\hat{\mathbf{X}}^{(t-1)}_{\mathrm{TF,d}}$ and $\hat{\mathbf{H}}^{(t-1)}_{\mathrm{DD}}$, and the signal after the DD-domain IC is given by}
\vspace{-2mm}
\begin{align}
\tilde{\mathbf{Y}}^{(t)}_{\mathrm{DD}}&[l',k'] 
=\mathbf{Y}_{\mathrm{DD}}[l',k']-\hat{\mathbf{I}}_{\mathrm{DD,d}}^{(t)}[l',k'] \notag \\
&=
h_{l',k'}\sqrt{N'}x_{\mathrm{pp}}+\Delta\mathbf{I}^{(t)}_{\mathrm{DD,d}}[l',k'] +\mathbf{W}_{\mathrm{DD}}[l',k'],
\label{DDIC}
\end{align}
\par\vspace{-2mm}\noindent
where $\Delta\mathbf{I}^{(t)}_{\mathrm{DD,d}}[l',k']={\mathbf{I}}_{\mathrm{DD,d}}[l',k']-\hat{\mathbf{I}}_{\mathrm{DD,d}}^{(t)}[l',k']$ denotes the residual interference at the $[l',k']$-th DD bin and $\hat{\mathbf{I}}_{\mathrm{DD,d}}^{{(t)}}[l',k']$ is the estimated interference signal, i.e., 
\vspace{-1mm}
\begin{equation}
\hat{\mathbf{I}}_{\mathrm{DD,d}}^{{(t)}}[l',k'] =
\sum_{l=0}^{l_{\rm max}}\sum_{k\in\mathcal D}\hat{h}^{(t-1)}_{l,k}\,\hat{f}^{(t-1)}_{l',k',l,k},\vspace{-1mm}
\label{YDDdhat}
\end{equation}
\textcolor{black}{where $\hat{f}^{(t-1)}_{l',k',l,k}$ is given by \eqref{fmi} with $\mathbf{X}_{\mathrm{TF,d}}$ replaced by $\hat{\mathbf{X}}^{(t-1)}_{\mathrm{TF,d}}$. It should be noted that, in the initial stage, IC is not applied, i.e., $\hat{\mathbf{I}}_{\mathrm{DD,d}}^{(0)}[l',k']=0$, $\tilde{\mathbf{Y}}^{(0)}_{\mathrm{DD}}[l',k'] = {\mathbf{Y}}_{\mathrm{DD}}[l',k']$ and $\Delta\mathbf{I}^{(0)}_{\mathrm{DD,d}}[l',k']=\mathbf{I}_{\mathrm{DD,d}}[l',k']$.}

    For $t\ge1$, let $\Delta h^{(t-1)}_{l',k'} = h_{l',k'} - \hat{h}^{(t-1)}_{l',k'}$ denote the CE error and $\Delta f^{(t-1)}_{l',k',l,k} = f_{l',k',l,k} - \hat{f}^{(t-1)}_{l',k',l,k}$ denote the estimation error of the interference component at the $[l',k']$-th DD bin in the $(t-1)$-th iteration. Then, the residual interference at the $[l',k']$-th DD bin is given by
\vspace{-2mm}
\begin{align}
\Delta\mathbf{I}^{(t)}_{\mathrm{DD,d}}&[l',k']=\sum_{l=0}^{l_{\rm max}}\sum_{k\in\mathcal D} h_{l,k}\,f_{l',k',l,k}-\hat{h}^{(t-1)}_{l,k}\,\hat{f}^{(t-1)}_{l',k',l,k}\notag\\[-2.8mm]
& \approx\sum_{l=0}^{l_{\rm max}}\sum_{k\in\mathcal D}\Delta h^{(t-1)}_{l,k}\,f_{l',k',l,k}+h_{l,k}\Delta f^{(t-1)}_{l',k',l,k}, \vspace{-1mm}
\end{align}
\par\vspace{-1mm}\noindent
where the approximation follows via neglecting the second-order error term 
$\Delta h^{(t-1)}_{l,k} \Delta f^{(t-1)}_{l',k',l,k}$. \textcolor{black}{Not surprisingly, the residual interference in the DD-domain is governed by the CE error and the TF-domain symbol detection error from the previous iteration. Furthermore, the mean and variance of the DD-domain residual interference are approximately given by
    \vspace{-1mm}\begin{equation}
    \mathbb{E}[\Delta\mathbf{I}^{(t)}_{\mathrm{DD,d}}[l',k']]\overset{(a)}{\approx} 0,\vspace{-2mm}
    \end{equation}
and
    \vspace{-1mm}\begin{align}
    \label{EIDDlk2}
    \mathbb{E}[|\Delta\mathbf{I}^{(t)}_{\mathrm{DD,d}}&[l',k']|^2]
    \overset{(b)}{\approx} \sum_{l=0}^{l_{\rm max}}\sum_{k\in\mathcal D} \Big( \mathbb{E} [|\Delta h^{(t-1)}_{l,k}|^2]\,\notag \\[-1mm]
    & \cdot \mathbb{E}[|f_{l',k',l,k}|^2]+ |h_{l,k}|^2\,\mathbb{E} [|\Delta f^{(t-1)}_{l',k',l,k}|^2 ]\Big), \vspace{-6mm}
    \end{align}
    respectively, where the approximations (a) and (b) follow from assuming that the CE errors are zero-mean, mutually uncorrelated, and uncorrelated with both the data symbols in the DD-domain and the channel coefficients. When $t=0$, $\mathbb{E}[|\Delta\mathbf{I}^{(0)}_{\mathrm{DD,d}}[l',k']|^2]$ is calculated by \eqref{IDDdlk}.} Let $\Delta {\mathbf{X}}^{(t-1)}_{\mathrm{TF,d}}[m',n']=\mathbf{X}_{\mathrm{TF,d}}[m',n']-\hat{\mathbf{X}}^{(t-1)}_{\mathrm{TF,d}}[m',n']$ denote the symbol detection error in the TF-domain. Since $\Delta{\mathbf{X}}^{(t)}_{\mathrm{TF,d}}[m',n']$ are approximately uncorrelated for $m'=0,\ldots,M-1$ and $n'=0,\ldots,N'-1$ $($as will be shown later in Section \ref{Eq}$)$, $\mathbb{E}\left[|\Delta f^{(t-1)}_{l',k',l,k}|^2\right]$ in \eqref{EIDDlk2} can be derived as \eqref{Edeltaf} at the top of the next
    page.
        \begin{figure*}[!t] 
        \centering
        \vspace*{-5mm}
    \begin{align}
    \mathbb{E}\left[|\Delta f^{(t-1)}_{l',k',l,k}|^2\right] =\begin{cases}  0, & \text{if } l'=l\\\frac{1}{MN}\sum_{n'=1}^{N'-1}\sum_{m'=0}^{M-1}\mathbb{E}[|\Delta {\mathbf{X}}^{(t-1)}_{\mathrm{TF,d}}[m',n']|^2], & \text{if } l' < l\\
    \frac{1}{MN}\sum_{n'=0}^{N'-1}\sum_{m'=0}^{M-1}\mathbb{E}[|\Delta {\mathbf{X}}^{(t-1)}_{\mathrm{TF,d}}[m',n']|^2], & \text{if } l' > l
    \end{cases}
    \label{Edeltaf}
        \end{align} 
        \par\vspace{-2mm}\noindent
            \hrulefill 
            \vspace{-6mm}
    \end{figure*}
\textcolor{black}{Combining \eqref{fmi2}, \eqref{EIDDlk2}, and \eqref{Edeltaf}, it can be observed that $\mathbb{E}[|\Delta\mathbf{I}^{(t)}_{\mathrm{DD,d}}[l',k']|^2]$ is independent of $k'$, i.e., within each delay tap, all Doppler bins share an identical residual interference power. Since \eqref{EIDDlk2} involves the true channel coefficients, which are unavailable, we approximate them using the channel estimates from the previous iteration. Thus, we have
\vspace{-1mm}
\begin{align}
    \label{Il}
    \mathbb{E}[|I^{(t)}_{l'}|^2]&
    = \sum_{l=0}^{l_{\rm max}}\sum_{k\in\mathcal D} \Big( \mathbb{E} [|\Delta h^{(t-1)}_{l,k}|^2]\,\mathbb{E}[|f_{l',k',l,k}|^2] \notag\\[-1.5mm]
    &+ |\hat{h}_{l,k}^{(t-1)}|^2\,\mathbb{E} [|\Delta f^{(t-1)}_{l',k',l,k}|^2 ]\Big)+N_0, \vspace{-4mm}
\end{align}
where $\mathbb{E}[|I^{(t)}_{l'}|^2]$ denotes the approximated residual interference and noise power at each DD bin corresponding to the $l'$-th delay tap, whose accuracy is validated through simulations in Section \ref{chap:simulation}. Besides, in the initial stage with $t=0$, since IC is not yet applied, as discussed in Section~\ref{gaussian}, the interference in the DD-domain caused by the TF-domain data symbols can be well approximated by a Gaussian distribution. Without loss of generality, the channel gains are assumed to be normalized such that $\sum_{i=1}^{P} \sigma_i^2 = 1$ \cite{yuan_data_aided}. Accordingly, the symbol interference and noise at each DD bin are jointly modeled as additive white Gaussian noise with an average power of $\mathbb{E}[|I^{(0)}_{l'}|^2] = N_0 + E_\mathrm{s}$.}

\vspace{-2.5mm}
\subsection{DD-domain Channel Estimation}
 \label{Section:DDCE}
\textcolor{black}{Based on the residual interference and noise power analysis above, as commonly done in OTFS \cite{Thre2,Thre1}, a low-complexity thresholding method is adopted for DD-domain CE:}
\vspace{-1mm}\begin{equation}
\hat{h}^{(t)}_{l',k'}=
\begin{cases}
\frac{\tilde{\mathbf{Y}}^{(t)}_{\mathrm{DD}}[l',k'] }{\sqrt{N'}x_{\mathrm{pp}}}, & |\tilde{\mathbf{Y}}^{(t)}_{\mathrm{DD}}[l',k']|\ge \Gamma^{(t)}_{l'}, \\
0, & |\tilde{\mathbf{Y}}^{(t)}_{\mathrm{DD}}[l',k'] |<\Gamma^{(t)}_{l'},
\end{cases}\vspace{-1mm}
\label{DDCE}
\end{equation}
with $\Gamma^{(t)}_{l'}=3\sqrt{\mathbb{E}[|I^{(t)}_{l'}|^2]}$\cite{yuan_data_aided}. And the corresponding CE error can be expressed as
\vspace{-1mm}\begin{equation}
\Delta{h}^{(t)}_{l',k'}=
\begin{cases}
\frac{\Delta\mathbf{I}^{(t)}_{\mathrm{DD,d}}[l',k'] +\mathbf{W}_{\mathrm{DD}}[l',k']}{\sqrt{N'}x_{\mathrm{pp}}}, & |\tilde{\mathbf{Y}}^{(t)}_{\mathrm{DD}}[l',k']|\ge \Gamma^{(t)}_{l'}, \\
h_{l',k'}, & |\tilde{\mathbf{Y}}^{(t)}_{\mathrm{DD}}[l',k'] |<\Gamma^{(t)}_{l'}.
\end{cases}\vspace{-1mm}
\end{equation}
\textcolor{black}{Subsequently, by replacing ${h}_{l',k'}$ in \eqref{EIDDlk2} with the most recent channel estimate $\hat{h}^{(t)}_{l',k'}$, and denoting the resulting quantity by $\mathbb{E}[|\tilde{I}^{(t)}_{l'}|^2]$, and let ${\mathbf{C}}^{(t)}_{\mathrm{DD}}[l',k']=\mathbb{E}[|\Delta{h}^{(t)}_{l',k'}|^2]$ for notational convenience, the CE error variance is then given by}
\vspace{-1mm}\begin{equation}
{\mathbf{C}}^{(t)}_{\mathrm{DD}}[l',k']=
\begin{cases}
\frac{\mathbb{E}[|\tilde{I}^{(t)}_{l’}|^2]}{N'E\mathrm{pp}}, & |\tilde{\mathbf{Y}}^{(t)}_{\mathrm{DD}}[l',k']|\ge \Gamma^{(t)}_{l'}, \\
|h_{l',k'}|^2, & |\tilde{\mathbf{Y}}^{(t)}_{\mathrm{DD}}[l',k']|<\Gamma^{(t)}_{l'}.
\end{cases}\vspace{-1mm}
\label{DDerror}
\end{equation}
Since the true value of $h_{l',k'}$ in \eqref{DDerror} is unavailable, the error variance for $|\tilde{\mathbf{Y}}^{(t)}_{\mathrm{DD}}[l',k']|<\Gamma^{(t)}_{l'}$ cannot be directly evaluated. \textcolor{black}{To promote the sparsity of the DD-domain channel, we set $\mathbb{E}[|\Delta{h}^{(t)}_{l',k'}|^2]=0$ when $|\tilde{\mathbf{Y}}^{(t)}_{\mathrm{DD}}[l',k']|<\Gamma^{(t)}_{l'}$ in practice. Besides, as a compromise, we set a large $P_{\max} > P$ to avoid potential path miss detection, i.e., when the paths number detected by thresholding exceeds $P_{\max}$, those ranked beyond $P_{\max}$ by path gain are treated as false and set to zero.}

\vspace{-2.5mm}
\subsection{\textcolor{black}{TF-domain Symbol Detection}}
 \label{Eq}
 \textcolor{black}{Based on the DD-domain CE results, the MMSE channel equalization is first performed to obtain TF-domain symbol estimates. Subsequently, a soft demodulation scheme is performed based on the constellation, yielding the posterior distribution of the TF-domain data symbols.}
 
Substituting the channel estimate from the $(t)$-th iteration into \eqref{eq:matrix_form} yields
\begin{equation}
\mathbf{y}^{(n')}_{\mathrm{TF,d}} =\hat{\mathbf{H}}_{\text{TF,d}}^{(t,n')} \mathbf{x}^{(n')}_{\mathrm{TF,d}} +\mathbf{i}_{\mathrm{TF,d}}^{(t,n')}+\mathbf{w}^{(n')}, \label{esti_yTF}\vspace{-2mm}
\end{equation}
\textcolor{black}{where $\mathbf{i}_{\mathrm{TF,d}}^{(t,n')}=\Delta{\mathbf{H}}_{\text{TF,d}}^{(t,n')} \mathbf{x}^{(n')}_{\mathrm{TF,d}}$ is the TF-domain interference caused by DD-domain CE errors}, $\Delta{\mathbf{H}}_{\text{TF,d}}^{(t,n')}={\mathbf{H}}_{\text{TF,d}}^{(t,n')}-\hat{\mathbf{H}}_{\text{TF,d}}^{(t,n')}\in\mathbb C^{M\times M}$, and the reconstructed TF-domain channel for the $n'$-th OFDM symbol in the $(t)$-th iteration is given by
\vspace{-2mm}\begin{equation}
\label{H_TF_hat}
\hat{\mathbf{H}}_{\text{TF,d}}^{(t,n')} \hspace{-1mm}
=\hspace{-1mm} \sum_{l=0}^{l_{\rm max}}\sum_{k\in\mathcal D} \hat{h}^{(t)}_{l',k'}e^{j 2\pi k\left(\frac{n'}{N'}+\frac{M_{\mathrm{CP}}+1-l'}{N'M_{\mathrm{De}}}\right) }\mathbf{F}_M\mathbf{\Delta}_M^{k'}\mathbf{\Pi}_M^{l'}\mathbf{F}_M^{\text{H}}.\vspace{-0mm}
\end{equation}
\textcolor{black}{In conventional TF-domain pilot and CE schemes, channel reconstruction is 
performed by interpolating the channel coefficients estimated on individual 
subcarriers (e.g., linear interpolation or ideal 2D low-pass filtering) 
\cite{COFDM2,DDaOFDM}. However, the rapidly varying TF-domain channel 
in high-mobility scenarios can introduce significant interpolation errors. In contrast, the proposed scheme directly estimates the delay-Doppler channel 
parameters, enabling more accurate TF-domain channel reconstruction than 
interpolation-based methods and characterizing the ICI channel structure, i.e., the 
off-diagonal entries in \eqref{H_TF_hat}. This improves the accuracy of the 
reconstructed TF-domain channel and enables ICI-aware equalization to overcome 
the performance limitation of conventional OFDM in high mobility channels}. 

In this work, since the TF-domain equalization is performed based on 
$\hat{\mathbf{H}}_{\text{TF,d}}^{(t,n')}$, we treat the interference
$\mathbf{i}_{\mathrm{TF,d}}^{(t,n')}$ as noise even though it contains a portion of the desired signal. Under the assumptions stated previously, we have $\mathbb{E}[\mathbf{i}_{\mathrm{TF,d}}^{(t,n')}[m']]=0$, $m'=0,1,...,M-1$, and
\vspace{-2mm}\begin{equation}
\mathbb{E}[\mathbf{i}_{\mathrm{TF,d}}^{(t,n')}(\mathbf{i}_{\mathrm{TF,d}}^{(t,n')})^\mathrm{H}]
=Es\sum_{l=0}^{l_{\rm max}}\sum_{k\in\mathcal D}\mathbb{E}[|\Delta{h}^{(t)}_{l',k'}|^2]\mathbf{I}_M.\vspace{-2mm}
\end{equation}

\subsubsection{Full-matrix MMSE Equalization}
\textcolor{black}{A full-matrix MMSE equalization is adopted to suppress ICI. The estimated TF-domain symbols are given by}
\begin{equation}
\tilde{\mathbf{x}}^{(t,n')}_{\mathrm{TF,d}} 
\hspace{-1mm}=\hspace{-1mm} E_\mathrm{s} (\hat{\mathbf{H}}_{\text{TF,d}}^{(t,n')})^{\text{H}}\hspace{-1mm}
\left(\hspace{-0.5mm} E_\mathrm{s}\hat{\mathbf{H}}_{\text{TF,d}}^{(t,n')}(\hat{\mathbf{H}}_{\text{TF,d}}^{(t,n')})^{\text{H}} \hspace{-0.5mm}+\hspace{-0.5mm} \eta^{(t)}\mathbf{I}_M \right)^{-1}\hspace{-1mm}\mathbf{y}_{\mathrm{TF,d}}^{(t,n')},\label{TFEq}
\vspace{-2mm}\end{equation}
with the estimation error covariance matrix
\vspace{-2mm}\begin{equation}
\tilde{\mathbf{C}}_{\mathrm{TF,d}}^{(t,n')}
=
\left(
\frac{1}{E_\mathrm{s}}\mathbf{I}_M
+ \frac{1}{\eta^{(t)}}(\hat{\mathbf{H}}_{\text{TF,d}}^{(t,n')})^{\text{H}} \hat{\mathbf{H}}_{\text{TF,d}}^{(t,n')}
\right)^{-1},\vspace{-2mm}
\label{TFerror}
\end{equation}
\textcolor{black}{where $\eta^{(t)}$ denotes the power of interference and noise for TF-domain equalization, given by}
\vspace{-2mm}\begin{equation}
\eta^{(t)}=E_\mathrm{s}\sum_{l=0}^{l_{\rm max}}\sum_{k\in\mathcal D}\mathbb{E}[|\Delta{h}^{(t)}_{l',k'}|^2]+N_0.\vspace{-2mm}
\end{equation}
In practice, the off-diagonal elements of ${\mathbf{C}}_{\mathrm{TF,d}}^{(t,n')}$ can often be neglected, as they typically have much smaller magnitudes compared to the diagonal elements \cite{cross_domain}\cite{cross_domain_FDE}. Therefore, the estimation errors 
$\Delta{\tilde{\mathbf{X}}}^{(t)}_{\mathrm{TF,d}}[m',n']$ 
can be treated as approximately uncorrelated 
for $m'=0,\ldots,M-1$ and $n'=0,\ldots,N'-1$. 
Accordingly, let $\tilde{\mathbf{C}}^{(t)}_{\mathrm{TF}}[m',n']=\mathbb{E}[|\Delta{\tilde{\mathbf{X}}}^{(t)}_{\mathrm{TF,d}}[m',n']|^2]$ for notational convenience, the symbol equalization error variance at the $[m',n']$-th TF resource element is given by
\vspace{-1mm}\begin{equation}
\tilde{\mathbf{C}}^{(t)}_{\mathrm{TF}}[m',n']=\tilde{\mathbf{C}}_{\mathrm{TF,d}}^{(t,n')}[m',m'].
\end{equation} 

\subsubsection{Single-tap MMSE Equalization}
\textcolor{black}{When the maximum Doppler is much smaller than the subcarrier space, a single-tap MMSE equalization can be applied to each TF resource element, thereby significantly reducing the computational complexity \cite{OFDMnoDoopler}. 
In this case, the estimated TF-domain symbol is given by \cite{cross_domain_FDE}}
\vspace{-1mm}
\begin{equation}
{\tilde{\mathbf{X}}^{(t)}_{\mathrm{TF,d}}[m',n']}=\frac{E_\mathrm{s}(\hat{\mathbf{H}}^{(t,n')}_{\mathrm{TF,d}}[m',m'])^{*}\mathbf{Y}_{\mathrm{TF}}[m',n']}{\sum\nolimits_{z'=0}^{M-1}E_\mathrm{s}|\hat{\mathbf{H}}^{(t,n')}_{\mathrm{TF,d}}[m',z']|^2+\eta^{(t)}}
\label{TFEqsingle}
\vspace{-1mm}
\end{equation} 
with the symbol equalization error variance given by
\vspace{-1mm}
\begin{equation}
\tilde{\mathbf{C}}^{(t)}_{\mathrm{TF}}[m',n'] 
=\frac{\sum_{\substack{z'=0 \\ z' \neq m'}}^{M-1}E_\mathrm{s}|\hat{\mathbf{H}}^{(t,n')}_{\mathrm{TF,d}}[m',z']|^2+\eta^{(t)}}{\sum_{z'=0}^{M-1}E_\mathrm{s}|\hat{\mathbf{H}}^{(t,n')}_{\mathrm{TF,d}}[m',z']|^2+\eta^{(t)}}.
\label{TFerrorsingle}
\vspace{-1mm}
\end{equation} 

		\begin{algorithm}[t]
				\caption{CD-IED Algorithm}
				\begin{algorithmic}[1]
					\State 
                    \textbf{Input:} $\mathbf{Y}_{\mathrm{DD}}$, $\mathbf{Y}_{\mathrm{TF}}$, $P_{\rm max}$, $t=0$, maximum iteration time $t_{\rm max}$ and stopping threshold $\varepsilon$. 
                    \Repeat
                    \State DD-domain IC: update $\tilde{\mathbf{Y}}^{(t)}_{\mathrm{DD}}$ by \eqref{DDIC}.
                    \State DD-domain CE: update $\hat{\mathbf{H}}^{(t)}_{\mathrm{DD}}$ and $\mathbf{C}^{(t)}_{\mathrm{DD}}$ by \eqref{DDCE} and  \eqref{DDerror}.
                    \State TF-domain equalization: update $\hat{\mathbf{H}}_{\text{TF,d}}^{(t,n')}$ by \eqref{H_TF_hat},   $\tilde{\mathbf{X}}^{(t)}_{\mathrm{TF,d}}$ by \eqref{TFEq} or \eqref{TFEqsingle}, and $\tilde{\mathbf{C}}^{(t)}_{\mathrm{TF,d}}$ by \eqref{TFerror} or \eqref{TFerrorsingle}.
                    \State Soft demodulation: update $\hat{\mathbf{X}}^{(t)}_{\mathrm{TF,d}}$ and $\mathbf{C}^{(t)}_{\mathrm{TF,d}}$ by \eqref{constelEq} and \eqref{constelerror}.
                    \State $t = t + 1$.
					\Until{$t = t_{\max}$ or $ ||\hat{\mathbf{X}}_{\mathrm{TF,d}}^{(t-1)}-\hat{\mathbf{X}}_{\mathrm{TF,d}}^{(t-2)}||_F \leq \varepsilon$}.
                    \State \textbf{Output:}   $\hat{\mathbf{H}}^{(t-1)}_{\mathrm{DD}}$, $\hat{\mathbf{X}}_{\mathrm{TF,d}}^{(t-1)}$.
                    \label{algorithm1}
				\end{algorithmic}
			\end{algorithm}

\subsubsection{\textcolor{black}{Soft Demodulation}}
\textcolor{black}{To further exploit constellation prior information of data symbols in the TF-domain, a probabilistic refinement is further performed \cite{cross_domain_FDE}, thereby facilitating subsequent iterative IC and CE. Specifically, the likelihood that the estimated symbol $\hat{\mathbf{X}}^{(t)}_{\mathrm{TF,d}}[m',n']$ corresponding to 
the constellation point $a_j \in \mathcal{A}$ is given by}
$P_{m',n',j}=\frac{\tilde{P}_{m',n',j}}{\sum_{k=1}^{|\mathcal{A}|}\tilde{P}_{m',n',k}}$,
with $\tilde{P}_{m',n',j}=\exp{(\frac{-|\tilde{\mathbf{X}}^{(t)}_{\mathrm{TF,d}}[m',n']-a_j |^2}{\tilde{\mathbf{C}}^{(t)}_{\mathrm{TF}}[m',n']} )}$. Then, the refined soft symbol estimate (i.e., the posterior mean) is obtained as
\vspace{-1mm}
\begin{equation}
\hat{\mathbf{X}}^{(t)}_{\mathrm{TF,d}}[m',n']=\sum\nolimits_{j=1}^{|\mathcal{A}|}P_{m',n',j}\,a_j,
\label{constelEq}
\vspace{-1mm}\end{equation} 
and the corresponding detection error variance is updated according to
\vspace{-1mm}\begin{align}
{\mathbf{C}}^{(t)}_{\mathrm{TF}}[m',n'] 
&= \mathbb{E}[|\Delta{\mathbf{X}}^{(t)}_{\mathrm{TF,d}}[m',n']|^2]\notag\\
&\hspace{-1mm}=\hspace{-1mm}\sum\nolimits_{j=1}^{|\mathcal{A}|}P_{m',n',j}\,|a_j|^2 \hspace{-1mm}-\hspace{-1mm}|\hat{\mathbf{X}}^{(t)}_{\mathrm{TF,d}}[m',n']|^2.
\label{constelerror}
\vspace{-1mm}
\end{align} 
\textcolor{black}{Subsequently, the TF-domain symbol detection results, together with the corresponding error variance, are fed back to the DD-domain for the next round of IC and CE.} 

\vspace{-2mm}
\subsection{Complexity Analysis}
\vspace{-1mm}
The overall algorithm is summarized in Algorithm 1. The computational complexity of the proposed algorithm is mainly dominated by DD-domain interference reconstruction, TF-domain channel reconstruction and equalization. \textcolor{black}{Specifically, in each iteration, the DD-domain interference reconstruction is mainly performed according to (35), with a computational complexity of $\mathcal{O}\!\left(\hat{P}(l_{\max}+1)(2k_{\max}+1)MN\right)$, where $\hat{P}$ denotes the number of identified paths in the DD-domain. The TF-domain channel reconstruction is computed based on (45), which requires a complexity of $\mathcal{O}(\hat{P}NM\log M)$. For TF-domain equalization, the conventional full-matrix MMSE equalization in (47) incurs a computational complexity of $\mathcal{O}(NM^{3})$, while the low-complexity single-tap MMSE equalization in (51) reduces the complexity significantly to $\mathcal{O}(NM)$. Although the proposed receiver adopts an iterative structure, it usually converges within only a few iterations (e.g., three iterations in our simulations in Section V) and the involved operations can be further accelerated by parallel and pipelined implementations, demonstrating its practical real-time feasibility.}

\vspace{-2mm}
\subsection{\textcolor{black}{Extension to Fractional Doppler Case}}
\vspace{-1mm}
\label{Sec:Frac}
The proposed framework can be readily extended to the fractional Doppler case by replacing the per-DD-bin-based signal processing in Section \ref{Section:DDIC} and \ref{Section:DDCE} with a per-delay-tap-based implementation, thereby yielding DD-domain channel estimates and the corresponding estimation errors under the virtual Doppler resolution, which are subsequently fed back to the TF-domain for symbol detection in Section~\ref{Eq}.

\textcolor{black}{Specifically, we formulate an on-grid SSR problem \cite{channelCS} based on a virtual representation defined over a virtual sampling grid in the Doppler domain. Let $R_{\nu}$ denote the virtual sampling factor\footnote{\textcolor{black}{The virtual sampling factor $R_{\nu}$ determines the virtual Doppler-domain grid resolution. Increasing $R_{\nu}$ provides a denser virtual Doppler grid, reduces grid mismatch error and improves DD-domain channel representation accuracy, at the cost of an increased SSR complexity. Therefore, $R_{\nu}$ should be selected to balance representation accuracy and computational complexity. 
Alternatively, existing off-grid/gridless compressed sensing methods \cite{off_grid,off_gird_yang} can be incorporated to formulate the DD domain CE as corresponding SSR problems and integrated into the proposed DD-a-OFDM framework.}}. Then, each Doppler shift $\nu_i$ can be mapped as $\nu_i = \frac{k_i^r}{R_{\nu} N' T}$, where $\frac{1}{R_{\nu} N' T}$ represents the virtual Doppler resolution and $k_i^r$ is the corresponding integer index.} Let $\mathbf{y}_{\mathrm{DD},l'} = [\mathbf{Y}_{\mathrm{DD}}[l',0], \dots, \mathbf{Y}_{\mathrm{DD}}[l',N'-1]]^\mathrm{T}$ denote the received signal corresponding to the $l'$-th delay tap, and $\mathbf{h}^{R_\nu}_{l'} = [h^{R_\nu}_{l',0},\dots, h^{R_\nu}_{l',k^{R_{\nu}}_{\max}}, h^{R_\nu}_{l',R_{\nu}N'-k^{R_{\nu}}_{\max}},... h^{R_\nu}_{l',R_{\nu}N'-1}]^\mathrm{T} \in \mathbb{C}^{(2k^{R_{\nu}}_{\max}+1) \times 1}$ denotes the high-resolution sparse channel vector associated with the $l'$-th delay tap, 
where $k^{R_{\nu}}_{\max}=R_{\nu}k_{\max}$, $h^{R_\nu}_{l',r}=\sum_{i=1}^{P} g_i\delta([l' - l_i]_M)\delta([r - \nu_i R_{\nu} N' T]_{R_{\nu} N'})$, $l'=0,1,...l_{\rm max}$ and $r\in \mathcal{D}^{R_{\nu}}$ with $\mathcal{D}^{R_{\nu}}=\{0,1,\ldots,k^{R_{\nu}}_{\max}\}\cup\{R_{\nu}N'-k^{R_{\nu}}_{\max},\ldots,R_{\nu}N'-1\}$. By combining \eqref{DDrela} and \eqref{Y_DD_I}, the input-output relationship for the $l'$-th delay tap in the DD-domain can be formulated as
\vspace{-2mm}
\begin{equation}
\mathbf{y}_{\mathrm{DD},l'}= x_\mathrm{pp}\boldsymbol{\Phi}^{R_\nu} \mathbf{h}^{R_\nu}_{l'} + \mathbf{i}_{\mathrm{DD},l'} + \mathbf{w}_{\mathrm{DD},l'},
\label{y_DD_l'_frac}
\vspace{-1mm}\end{equation}
with $\mathbf{w}_{\mathrm{DD},l'}=[\mathbf{W}_{\mathrm{DD}}[l',0],\dots, \mathbf{W}_{\mathrm{DD}}[l',N'-1]]^\mathrm{T}$. In \eqref{y_DD_l'_frac}, $\boldsymbol{\Phi}^{R_\nu} \in \mathbb{C}^{N' \times (2k^{R_{\nu}}_{\max}+1)}$ denotes the measurement matrix, with $\boldsymbol{\Phi}^{R_\nu}[a,q] =\frac{1}{\sqrt{N'}} \sum_{n'=0}^{N'-1} e^{-j 2 \pi n' \frac{a - \frac{q'}{R_{\nu}}}{N'}}$, where 
\vspace{-2mm}
\begin{align}
\label{q'}
q' &= 
\begin{cases}
q, & \text{if } q\leq R_{\nu}k_{\max},\\
N'R_{\nu}+q-{2k^{R_{\nu}}_{\max}}-1, & \text{if } q > R_{\nu}k_{\max}.
\end{cases}
\vspace{-3mm}
\end{align}
And $\mathbf{i}_{\mathrm{DD},l'} \in \mathbb{C}^{N' \times 1}$ denotes the DD-domain interference vector corresponding to the $l'$-th delay tap, with $\mathbf{i}_{\mathrm{DD},l'}[k'] = \mathbf{I}_{\mathrm{DD,d}}[l', k']$. In order to represent the interference terms under the virtual Doppler grid, $\mathbf{i}_{\mathrm{DD},l'}$ can be rewritten as
\vspace{-2mm}\begin{equation}
\mathbf{i}_{\mathrm{DD},l'}=\sum\nolimits_{\substack{l=0, l \neq l'}}^{l_{\max}} \mathbf{D}_{l',l} \mathbf{h}^{R_\nu}_{l},
\label{iDDl}
\vspace{-2mm}\end{equation}
with $\mathbf{D}_{l',l}= [\mathbf{d}_{l',0,l},\mathbf{d}_{l',1,l},
\cdots ,\mathbf{d}_{l',N'-1,l}]^\mathrm{T}
\in \mathbb{C}^{N' \times (2k^{R_{\nu}}_{\max}+1)}$, where
\vspace{-2mm}
\begin{align}
\mathbf{d}_{l',k,l} &= 
\begin{cases}
\mathbf{Q}_{l',l} \check{\mathbf{F}}^k_N \check{{\mathbf{X}}}_{\mathrm{TF},d}^\mathrm{T} \mathbf{z}_{l',l}, & \text{if } l '< l\\
\mathbf{Q}_{l',l} \mathbf{F}^k_N \mathbf{X}_{\mathrm{TF},d}^\mathrm{T} \mathbf{z}_{l',l}, & \text{if } l' \ge l
\end{cases}
\end{align}
\par\vspace{-2mm}\noindent
is a data-dependent while channel-independent vector constructed from the TF-domain data symbols. 
Here, $\mathbf{Q}_{l',l}\in\mathbb{C}^{(2k^{R_{\nu}}_{\max}+1)\times(2k^{R_{\nu}}_{\max}+1)}$ is a diagonal matrix with $\mathbf{Q}_{l',l}[q,q]=e^{j2\pi \frac{(l'-l)q'}{M_{\mathrm{De}}N'R_{\nu}}}$, and $\mathbf{F}_N^k\in\mathbb{C}^{(2k^{R_{\nu}}_{\max}+1)\times N'}$ with $\mathbf{F}_N^k[q,a]=e^{j2\pi \frac{a(q'-kR_{\nu})}{N'R_{\nu}}}$, where $q'$ follows the same definition in \eqref{q'}.
$\mathbf{z}_{l,l'}\in \mathbb{C}^{M \times 1}$ is given by
\vspace{-2mm}
\begin{align}
\mathbf{z}_{l',l}[m] &= 
\begin{cases}
e^{j2\pi \frac{m[M+l'-l]_M}{M}}, & \text{if } l' < l,\\
e^{j2\pi \frac{m[l'-l-M_{CP}-1]_M}{M}}, & \text{if } l' \ge l.
\end{cases}
\end{align}
\par\vspace{-2mm}\noindent
$\check{\mathbf{F}}^k_N \in \mathbb{C}^{(2k^{R_{\nu}}_{\max}+1)\times (N'-1)}$ and ${\check{\mathbf{X}}}_{\mathrm{TF},d} \in \mathbb{C}^{M \times (N'-1)}$ denote the matrices obtained by removing the first column of $\mathbf{F}^k_N$ and $\mathbf{X}_{\mathrm{TF},d}$, respectively. It can be observed that \eqref{y_DD_l'_frac} decouples the data symbols and channel, similar to \eqref{eq:DD_IO_effective} and \eqref{YDDd}, which enables iterative IC and CE procedures similar to those presented in Section \ref{Section:DDIC} and \ref{Section:DDCE}. Specifically, at the $(t)$-th iteration, DD-domain IC is performed on a per-delay-tap basis, i.e.,
\vspace{-1mm}
\begin{subequations}
\begin{align}
\tilde{\mathbf{y}}_{\mathrm{DD},l'}^{(t)} &=\mathbf{y}_{\mathrm{DD},l'} -\hat{\mathbf{i}}_{\mathrm{DD},l'}^{(t)}
\\&=x_\mathrm{pp}\boldsymbol{\Phi}^{R_\nu} \mathbf{h}^{R_\nu}_{l'} +\Delta{\mathbf{i}}_{\mathrm{DD},l'}^{(t)}  +  \mathbf{w}_{\mathrm{DD},l'}, \label{SSR_frac}
\vspace{-2mm}
\end{align}
\end{subequations}
where $\hat{\mathbf{i}}_{\mathrm{DD},l'}^{(t)}$ denotes the estimated interference vector, which is given by \eqref{iDDl} with $\mathbf{X}_{\mathrm{TF,d}}$ and $\mathbf{h}^{R_\nu}_{l'}$ replaced by $\hat{\mathbf{X}}^{(t-1)}_{\mathrm{TF,d}}$ and $\hat{\mathbf{h}}^{R_\nu,(t-1)}_{l'}$ from the previous iteration, respectively. And $\Delta{\mathbf{i}}_{\mathrm{DD},l'}^{(t)} =\mathbf{i}_{\mathrm{DD},l'}-\hat{\mathbf{i}}_{\mathrm{DD},l'}^{(t)}\approx \sum\nolimits_{\substack{l=0,l\neq l'}}^{l_{\max}} \Delta{\mathbf{D}}_{l',l} ^{(t-1)}\mathbf{h}^{R_\nu}_{l}+\mathbf{D}_{l',l}\Delta{\mathbf{h}}^{R_\nu,(t-1)}_{l}$ denotes the residual interference vector of the $l'$-th delay tap.  $\Delta \mathbf{h}^{R_\nu,(t-1)}_{l} = \mathbf{h}^{R_\nu}_{l} - \hat{\mathbf{h}}^{R_\nu,(t-1)}_{l}$ denotes the CE error under the high Doppler resolution representation and $\Delta \mathbf{D}_{l',l}^{(t-1)} = \mathbf{D}_{l',l} - \hat{\mathbf{D}}_{l,l'}^{(t-1)}$ represents the estimation error of the constructed data-dependent matrix. \textcolor{black}{Based on the similar derivations in Sections IV-A and IV-B, the channel estimates 
$\hat{\mathbf{h}}^{R_\nu,(t)}_{l'}$ and the covariance matrix of the associated estimation 
error $\Delta\mathbf{h}^{R_\nu,(t)}_{l'}$ can be obtained by solving the SSR problem in 
\eqref{SSR_frac}, including LASSO \cite{LASSO}, orthogonal matching pursuit (OMP) 
\cite{OMP_CE}, and sparse Bayesian learning (SBL) \cite{SBL,off_grid,off_gird_yang}.} Due to space limitations, the details are omitted here. Based on these results, a cross-domain iterative framework for IC, CE and symbol detection can be established, as illustrated in Fig.~\ref{Iteration}.

	    \begin{table*}[b]
            \centering 
            \small 
            \vspace*{-3mm}
            \caption{Simulation parameters}\vspace*{-2mm}
                \begin{tabular}{c|c|c|c|c|c}
                    \hline 
                    Parameter & Value & Parameter & Value & Parameter & Value \\
                    \hline 
                    Carrier frequency $f_c$ (GHz) & 4 &No. of subcarriers ($M$) & 32 & Maximum delay spread $\tau_{\max}$ ($\mu$s)&  5.2\\Subcarrier spacing $\Delta f$ (kHz) & 30 & No. of time slots ($N$) & 32 & Maximum delay index $l_{\max}$ & 5 \\No. of Path ($P$) & 4  & $N'$ & 32 &Speed (km/h) & 300/600    \\ $P_{\rm max}$ & 6 & $t_{\rm max}$& 5 &Maximum Doppler index $k_{\max}$ & 1/2 \\
                    \hline 
                \end{tabular}
            \label{Simulation Parameters}
            \vspace*{-7mm}
        \end{table*}

\begin{figure*}[t]
\vspace{-7mm}
\centering
\begin{subfigure}{0.24\linewidth}
\centering
\includegraphics[width=\linewidth]{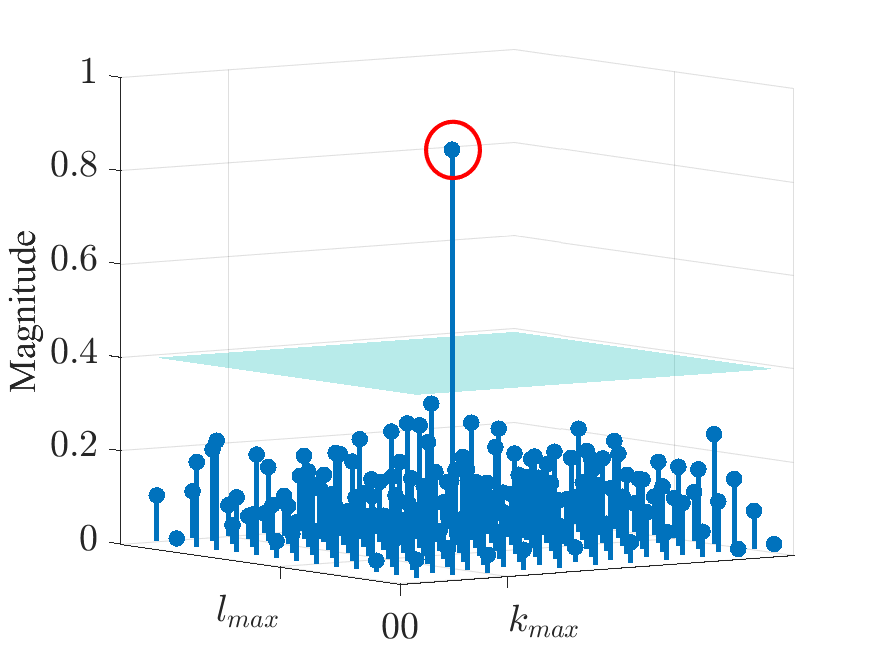}\vspace{-2mm}
\caption{CE with $\tilde{\mathbf{Y}}^{(0)}_{\mathrm{DD}}$}
\end{subfigure}
\hfill
\begin{subfigure}{0.24\linewidth}
\centering
\includegraphics[width=\linewidth]{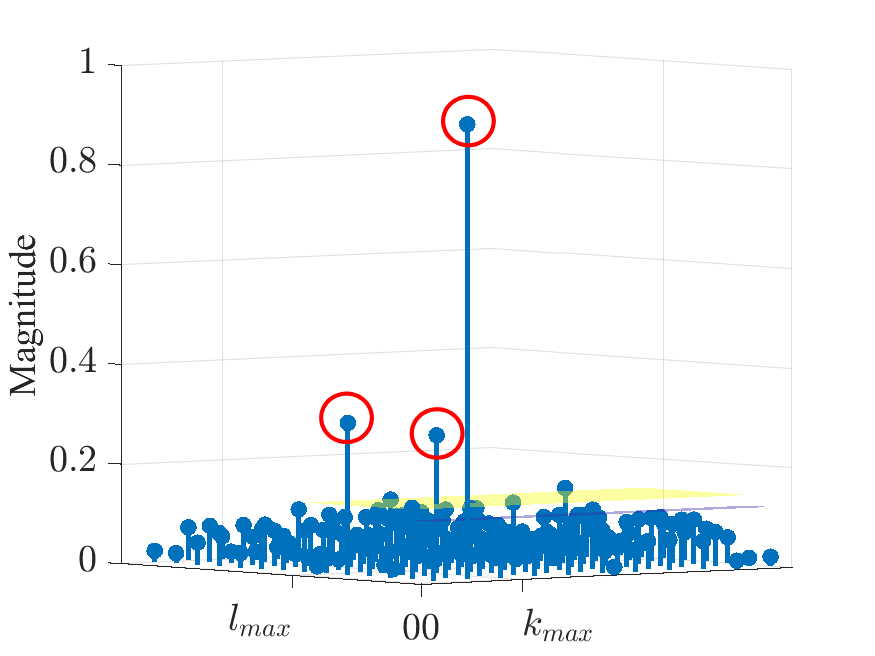}\vspace{-2mm}
\caption{CE with $\tilde{\mathbf{Y}}^{(1)}_{\mathrm{DD}}$}
\end{subfigure}
\hfill
\begin{subfigure}{0.24\linewidth}
\centering
\includegraphics[width=\linewidth]{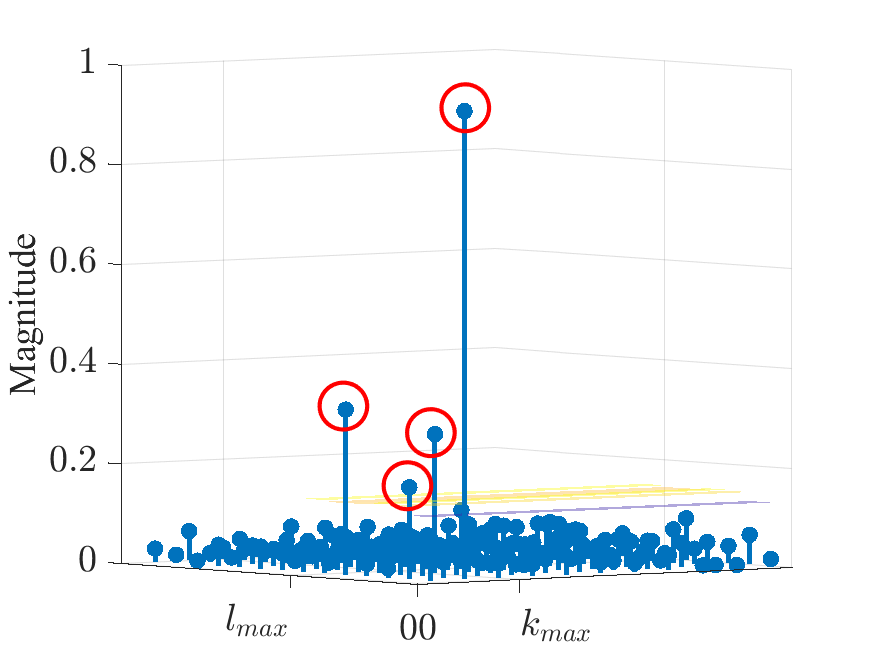}\vspace{-2mm}
\caption{CE with $\tilde{\mathbf{Y}}^{(3)}_{\mathrm{DD}}$}
\end{subfigure}
\hfill
\begin{subfigure}{0.24\linewidth}
\centering
\includegraphics[width=\linewidth]{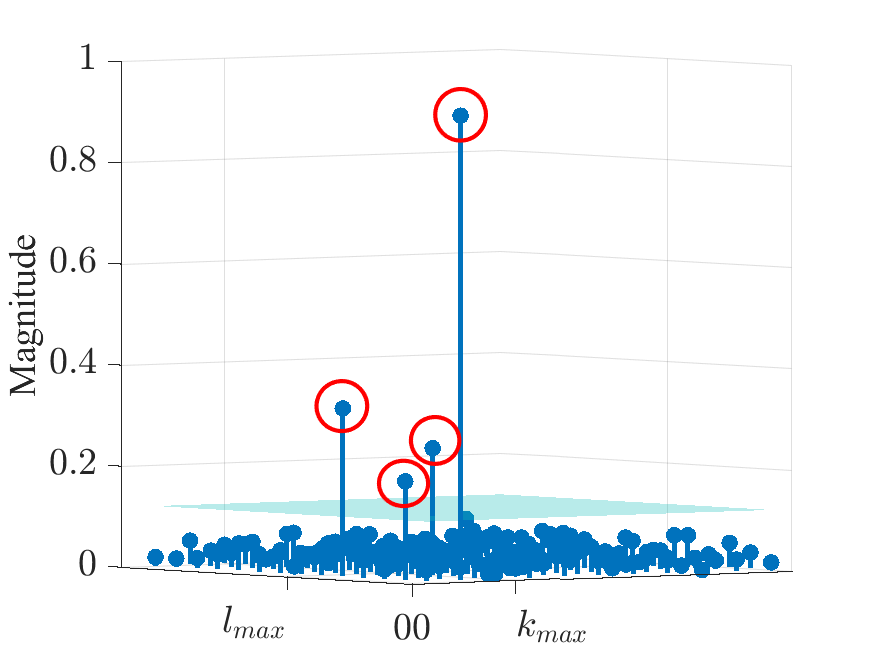}\vspace{-2mm}
\caption{CE with genie-aided IC}
\end{subfigure}
\vspace{-2mm}
\caption{\textcolor{black}{Example of iterative DD-domain CE. The channel coefficients are 
$[0.2427+0.2476i,\;0.3712-0.4061i,\;-0.0166-0.5722i,\;0.2182+0.4492i]$, 
with delay taps $[0,1,2,5]$ and Doppler taps $[2,1,1,1]$. 
The plane in each figure denotes the detection threshold and the red solid circles indicate the correctly detected DD channel taps.
}}
\label{tureCE}\vspace{-5mm}
\end{figure*}

\vspace{-1mm}
	\section{NUMERICAL RESULTS}
	\label{chap:simulation}
	In this section, simulations are conducted to verify the effectiveness of the proposed DD-a-OFDM transceiver architecture and the CD-IED algorithm.
    \subsection{Parameter Setting and Baseline Schemes}
	All the parameters are summarized in $\textbf{Table \ref{Simulation Parameters}}$. The integer delay index of each path is uniformly distributed over $[0, l_{\max}]$. Unless otherwise specified, the integer Doppler is considered in the simulations, where the Doppler indices are uniformly distributed within $[-k_{\max}, k_{\max}]$. Otherwise, the fractional Doppler case described in Section \ref{Sec:Frac} is considered, with an oversampling factor of $R_{\nu} = 10$. The path gain decays exponentially with respect to the delay. The signal-to-noise ratio (SNR) is defined by $\mathrm{SNR}=\frac{E_\mathrm{s}}{N_0}$. \textcolor{black}{Unless otherwise specified, we adopt a practical PP power design criterion rather than theoretically optimizing the PDR, since the CE, IC, and symbol equalization processes are strongly coupled in the proposed iterative receiver, making conventional normalized mean square error (NMSE)- or signal-to-interference-plus-noise ratio (SINR)-based PDR optimization non-trivial \cite{PA1,PA2}. Specifically, to improve CE reliability while avoiding high PAPR and facilitating practical implementation, the PP power is set according to the peak power of the transmitted time-domain OFDM signal, i.e., $E_{\mathrm{pp}}=\max|\mathbf{x}_{\mathrm{T,d}}[nM_{\mathrm{De}}+l]|^2$, $\forall n,l$.}
        
    In the simulations, three baseline schemes are considered for performance comparison:\\
    1) \textit{\textbf{Ideal CE in the DD-domain:}} In this scheme, perfect channel state information (CSI) in the DD-domain is assumed to be available at the receiver. This idealized assumption provides an upper bound on the performance of the considered system. The corresponding results are labeled as ‘Ideal’.\\
   2) \textit{\textbf{Genie-aided CE in the DD-domain:}} In this scheme, the interference in the DD-domain is assumed to be perfectly canceled with the aid of genie information, such that only additive noise remains. CE is then performed once according to \eqref{DDCE} with $\tilde{\mathbf{Y}}^{(0)}_{\mathrm{DD}}[l',k']
    = \mathbf{Y}_{\mathrm{DD}}[l',k'] - \mathbf{Y}_{\mathrm{DD,d}}[l',k']$
    and $\Gamma^{(1)}_{l'} = 3\sqrt{N_0}$. This scheme serves as a reference for CE performance under ideal IC. The corresponding results are labeled as ‘Genie-aided’.\\ 
    3) \textit{\textbf{Classical OFDM Structure:}} \textcolor{black}{In this scheme, pilot symbols occupy all subcarriers of the 11-$th$ and 22-$nd$ OFDM symbols \cite{COFDM2}, and the power of each pilot symbol is set to $E_{\mathrm{pp}}$.} \textcolor{black}{At the receiver, LS-based CE is first performed at the pilot locations in the TF-domain, followed by linear interpolation to obtain frequency-domain channel coefficients other than the 11-$th$ and 22-$nd$ OFDM symbols, based on which the single-tap equalization is applied. Note that this scheme can only recover per-subcarrier channel coefficients and does not explicitly capture the ICI channel as our proposed scheme. As a result, the full $M \times M$ frequency domain effective channel matrix for each OFDM symbol in \eqref{H_TF_hat} is not available, making it impossible to form the full-matrix MMSE equalizers.} The corresponding results are labeled as ‘Classical OFDM’.

\vspace{1mm}
	\subsection{Simulation Results}

\textcolor{black}{Fig. \ref{tureCE} illustrates iterative DD-domain CE result with PDR=3dB and SNR=10dB, where the full-matrix MMSE equalization is adopted. In Fig. \ref{tureCE}(a), we can observe that in the initial stage, the overall interference level in the DD-domain is relatively high, which submerges some DD-domain channel taps. As a result, only the strongest path is identified. In Fig. \ref{tureCE}(b), owing to TF-domain detection and DD-domain IC, the residual interference in the DD-domain is significantly reduced. Consequently, the CE threshold for each delay tap decreases, allowing the second and the third path to be detected. As the iteration proceeds, the residual interference level and the CE threshold for each delay tap continue to decrease, enabling more paths to be successfully identified. Morever, the CE performance approaches that of the genie-aided IC case in Fig. \ref{tureCE}(d) with $t=3$, demonstrating the effectiveness of the CD-IED algorithm.}

\begin{figure*}[t]
\centering
 \vspace{-7mm}
\begin{subfigure}{0.24\linewidth}
\centering
\includegraphics[width=\linewidth]{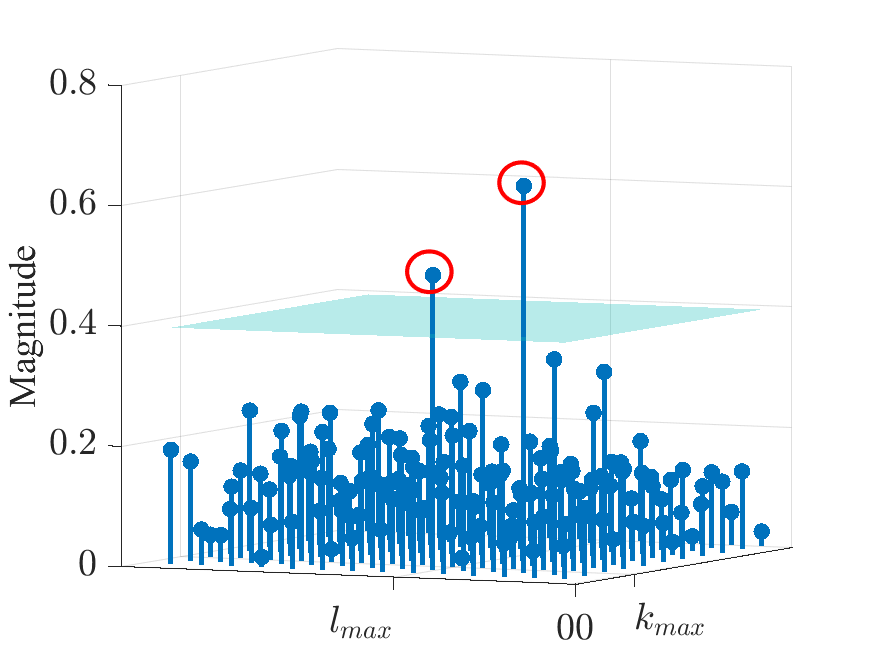}\vspace{-2mm}
\caption{CE with $\tilde{\mathbf{Y}}^{(0)}_{\mathrm{DD}}$}
\end{subfigure}
\hfill
\begin{subfigure}{0.24\linewidth}
\centering
\includegraphics[width=\linewidth]{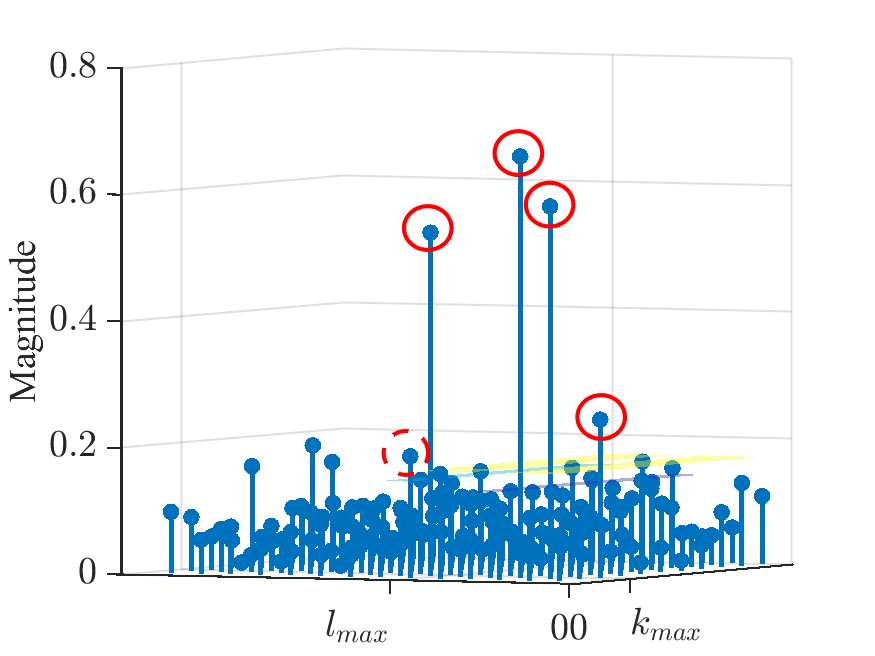}\vspace{-2mm}
\caption{CE with $\tilde{\mathbf{Y}}^{(1)}_{\mathrm{DD}}$}
\end{subfigure}
\hfill
\begin{subfigure}{0.24\linewidth}
\centering
\includegraphics[width=\linewidth]{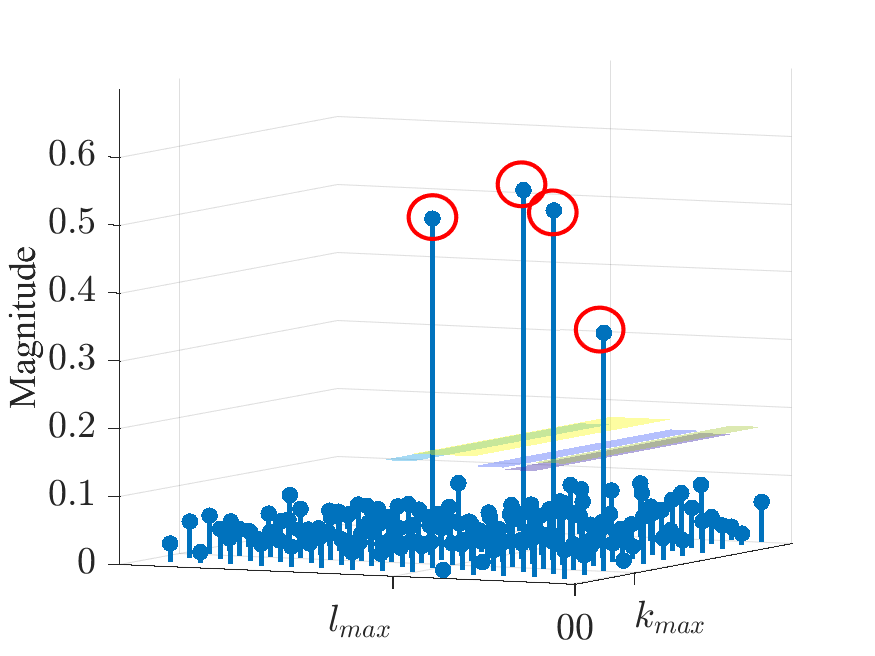}\vspace{-2mm}
\caption{CE with $\tilde{\mathbf{Y}}^{(3)}_{\mathrm{DD}}$}
\end{subfigure}
\hfill
\begin{subfigure}{0.24\linewidth}
\centering
\includegraphics[width=\linewidth]{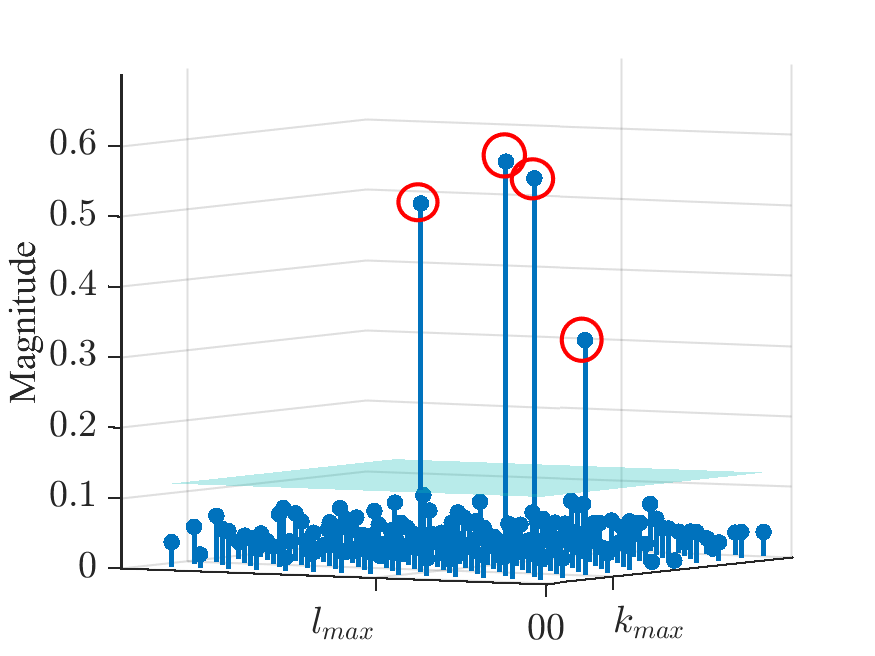}\vspace{-2mm}
\caption{CE with genie-aided IC}
\end{subfigure}\vspace{-1mm}
\caption{\textcolor{black}{Example of iterative DD-domain CE. The channel setup is the same as Fig. \ref{tureCE}.  
The plane denotes the detection threshold level. 
The red solid circles indicate the correctly detected channel taps, 
while the red dashed circles indicate fake channel taps caused by high-power noise and interference. 
}}
\label{falseCE}\vspace{-7mm}
\end{figure*}

Fig. \ref{falseCE} illustrates another example of iterative DD-domain CE with false channel tap detections. \textcolor{black}{Similar to Fig. \ref{tureCE}, as the iteration proceeds, the residual interference level gradually decreases and the number of detected DD-domain channel taps correspondingly increases. It is worth noting that in Fig. \ref{falseCE}(b), one fake channel tap is falsely detected due to high-power residual interference and noise. However, since the power of the fake channel tap primarily stems from residual interference and noise and it is much smaller than that of the correctly detected paths, the fake channel taps can be effectively suppressed and removed in subsequent iterations through more accurate interference reconstruction and cancellation. Consequently, a high CE accuracy can still be achieved, demonstrating the robustness of the proposed CD-IED algorithm.}

\begin{figure}[t]
    \centering\vspace{-2mm}
    \includegraphics[width=3in]{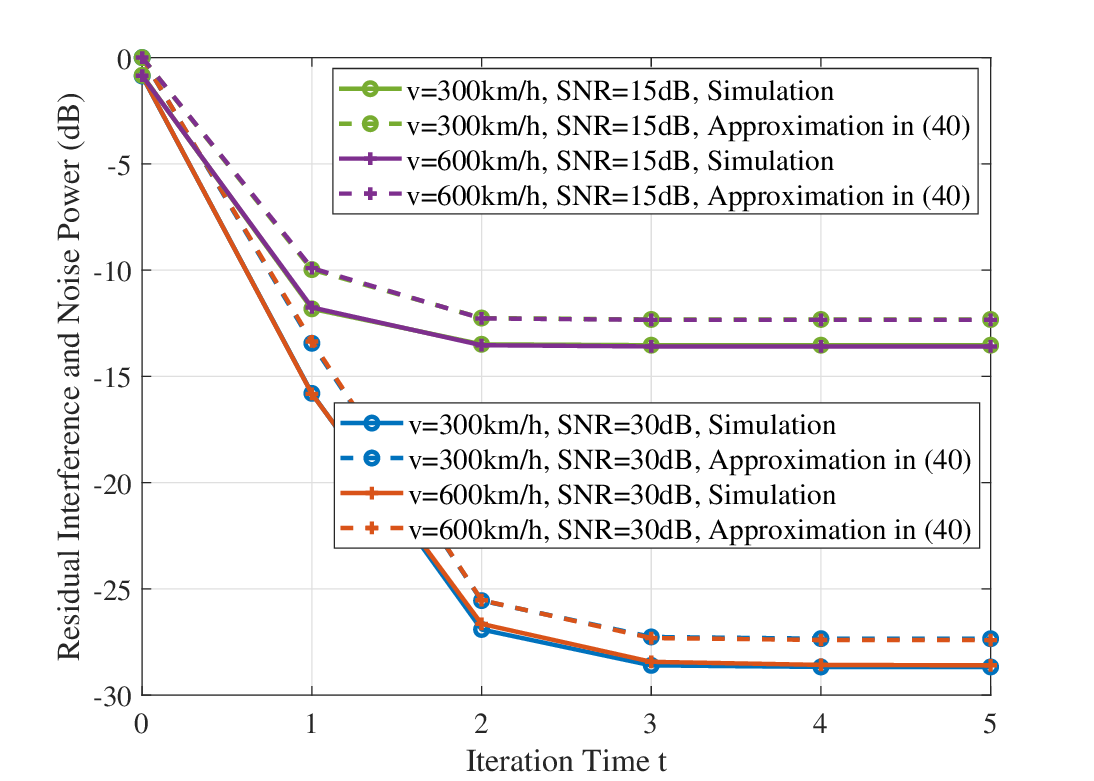}\vspace{-2mm} 
    \caption{DD-domain residual interference and noise power with different iteration time}
    \label{DD_I}\vspace{-4mm}
\end{figure}

Fig. \ref{DD_I} evaluates the approximation error of the residual interference and noise power in the DD-domain analysis in \eqref{Il} over iterations with full-matrix MMSE equalization. It can be observed that, for both $\mathrm{SNR}=15$ dB and $30$ dB, the residual interference-plus-noise power decreases rapidly as the iteration proceeds and remains robust against different velocities. Moreover, only a small discrepancy is observed between the analytical results in \eqref{Il} and the simulation results, which verifies the accuracy of the proposed residual interference-plus-noise power analysis after IC. This, in turn, facilitates more accurate DD-domain CE in subsequent iterations, thereby forming a beneficial iterative refinement process.

\begin{figure}[t]
    \centering\vspace{-2mm}
    \includegraphics[width=3in]{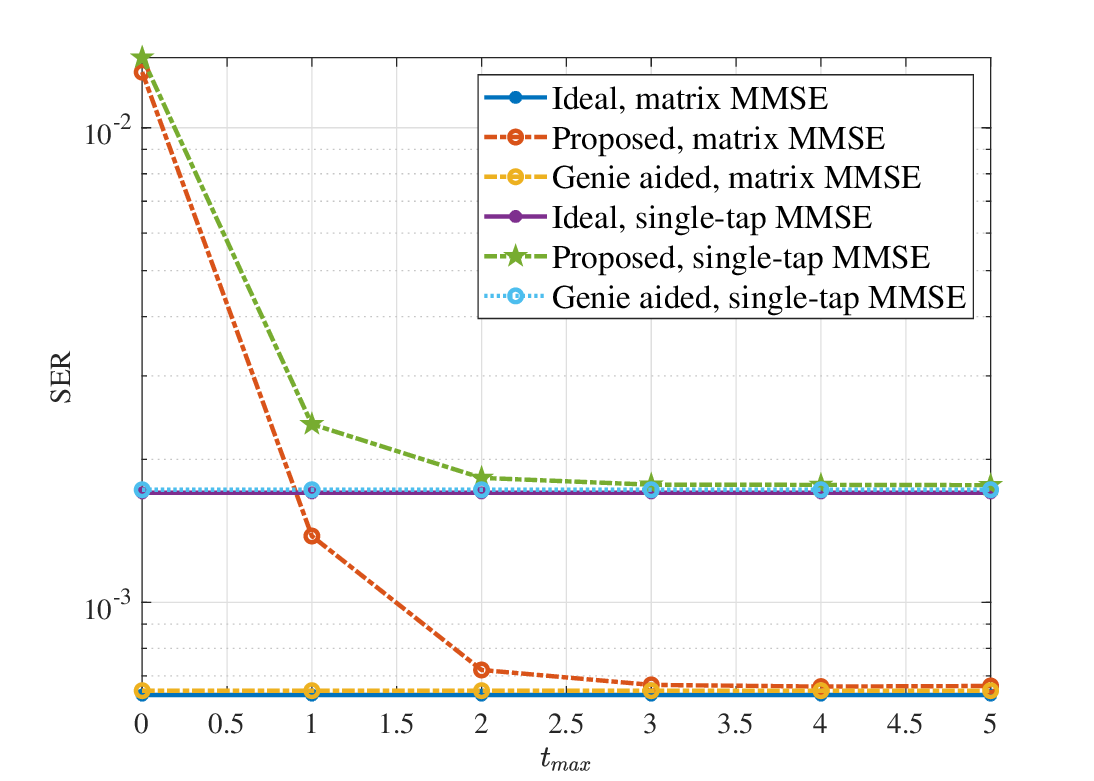}\vspace{-2mm}  
    \caption{SER performance with different maximum iteration number}
    \label{maxiter}\vspace{-7mm}
\end{figure}

Fig. \ref{maxiter} illustrates the symbol error rate (SER) performance under different maximum iteration numbers $t_{\rm max}$ \textcolor{black}{with SNR=30dB and speed of 300km/h. It can be observed that, for both single-tap MMSE and full-matrix MMSE equalization, the proposed CD-IED algorithm converges within three iterations, with detection performance approaching that of the ideal CSI and Genie-aided cases.} \textcolor{black}{This fast convergence mainly stems from the pulsone pilot design and the ordered path detection and estimation mechanism. Specifically, unlike the received data symbols that experience energy spreading and behave approximately as Gaussian noise in the DD-domain, the received time-domain PP sequence achieves 
coherent combining in the DD-domain, thereby improving the 
pilot-to-interference power ratio and enabling accurate channel estimation. More importantly, as illustrated in Fig. 6, the proposed CD-IED scheme naturally exhibits an ordered path detection and estimation strategy, where the paths are gradually detected according to their energy levels rather than being jointly estimated. Specifically, in the early iterations, the stronger paths are detected and estimated first, which captures the dominant channel components and improves the accuracy of TF-domain channel reconstruction and symbol detection. The resulting more accurate interference reconstruction and cancellation rapidly suppresses the dominant DD-domain interference, thereby facilitating the reliable detection of weaker paths in subsequent iterations and accelerating the convergence of the overall iterative process.}

\begin{figure}[t]
    \centering
    \includegraphics[width=3in]{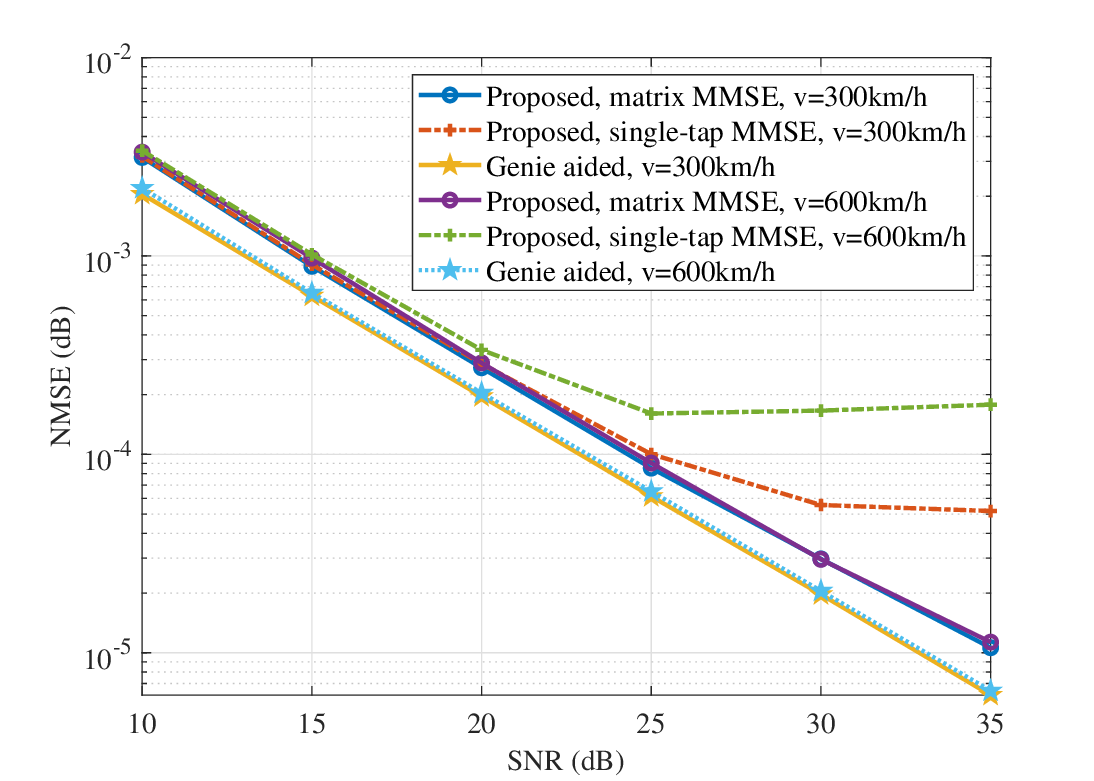}\vspace{-2mm}  
    \caption{NMSE performance of TF-domain channel estimation with different SNRs.}
    \label{NMSE}\vspace{2mm}
\end{figure}

Fig. \ref{NMSE} illustrates the normalized NMSE performance \textcolor{black}{of the TF-domain CE versus SNR under both speeds of 300km/h and 600km/h.} The NMSE is defined as
\vspace{-3mm}
\begin{align}
\mathrm{NMSE}
=\frac{1}{N'}\sum_{n'=0}^{N'-1}
\frac{\left\| \hat{\mathbf{H}}_{\mathrm{TF}}^{(n')} -{\mathbf{H}}_{\mathrm{TF}}^{(n')}\right\|_F^2}
{\left\|{\mathbf{H}}_{\mathrm{TF}}^{(n')} \right\|_F^2}.
\vspace{-4mm}
\end{align}
\textcolor{black}{We can observe that with increasing SNR, NMSE decreases with either single-tap MMSE or full-matrix MMSE equalization. In fact, when the full-matrix MMSE equalizer is employed, the ICI induced by high mobility can be effectively suppressed, enabling more accurate TF-domain detection and facilitating DD-domain IC and CE. As a result, the achieved NMSE performance approaches that of the Genie-aided case. In contrast, the single-tap MMSE equalizer cannot mitigate the severe ICI, leading to large detection errors in the TF-domain, which further degrade CE accuracy in both the DD and TF-domains, particularly at 600km/h.}

\begin{figure}[t]
    \centering\vspace{-3mm}
    \includegraphics[width=3in]{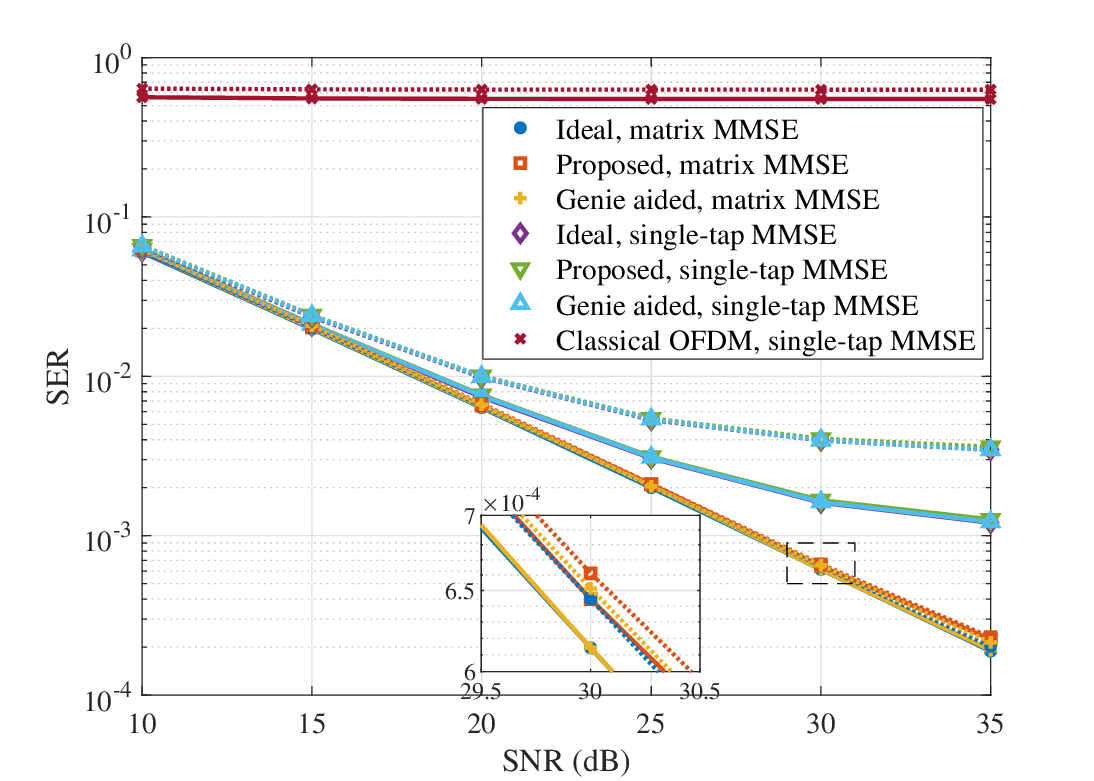}\vspace{-2mm}  
    \caption{SER performance in the integer Doppler case. The solid and dotted lines represent velocities of 300 and 600 km/h.}
    \label{SERInt}\vspace{-7mm}
\end{figure}

\begin{figure}[t]
    \centering
    \includegraphics[width=3in]{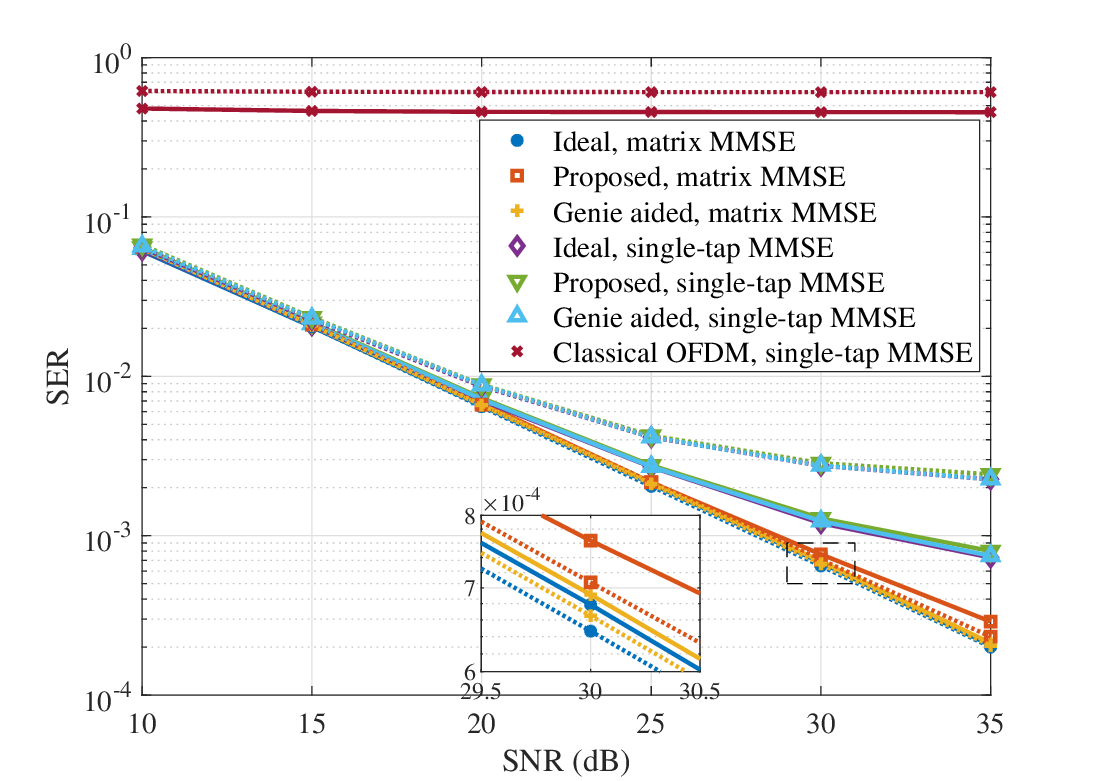}
    \vspace{-1.5mm}
    \caption{SER performance in the fractional Doppler case. The solid and dotted lines represent velocities of 300 and 600 km/h.}
    \label{SERFrac}\vspace{-6mm}
\end{figure}

Fig. \ref{SERInt} illustrate the SER performance versus SNR at speeds of 300km/h and 600km/h in the integer Doppler case. For both full-matrix and single-tap MMSE equalizers, the SER performance achieves that of the corresponding ideal CSI and genie-aided cases, and significantly outperforms classical OFDM transceiver scheme. In fact, the proposed pilot design and CD-IED algorithm can exploit both the DD-domain channel properties and the TF-domain channel reconstruction model in \eqref{H_TF_hat} for accurate CE and symbol detection. Nevertheless, a significant performance gap remains between the single-tap and full-matrix MMSE equalizers, and this gap becomes more pronounced as the user velocity increases. This is because the full-matrix equalizer is more effective at suppressing ICI induced by high mobility. Fig. \ref{SERFrac} further illustrates the SER performance in the fractional Doppler channel under the same settings as Fig. \ref{SERInt}. We observe almost no performance degradation when extending the proposed scheme from an integer Doppler channel to a fractional Doppler channel. In fact, owing to the coherent combination of the proposed PP sequence from the time-domain to the DD-domain, a simple on-grid SSR algorithm can correctly identify the channel Doppler parameters.

\begin{figure}[t]
    \centering
    \includegraphics[width=3in]{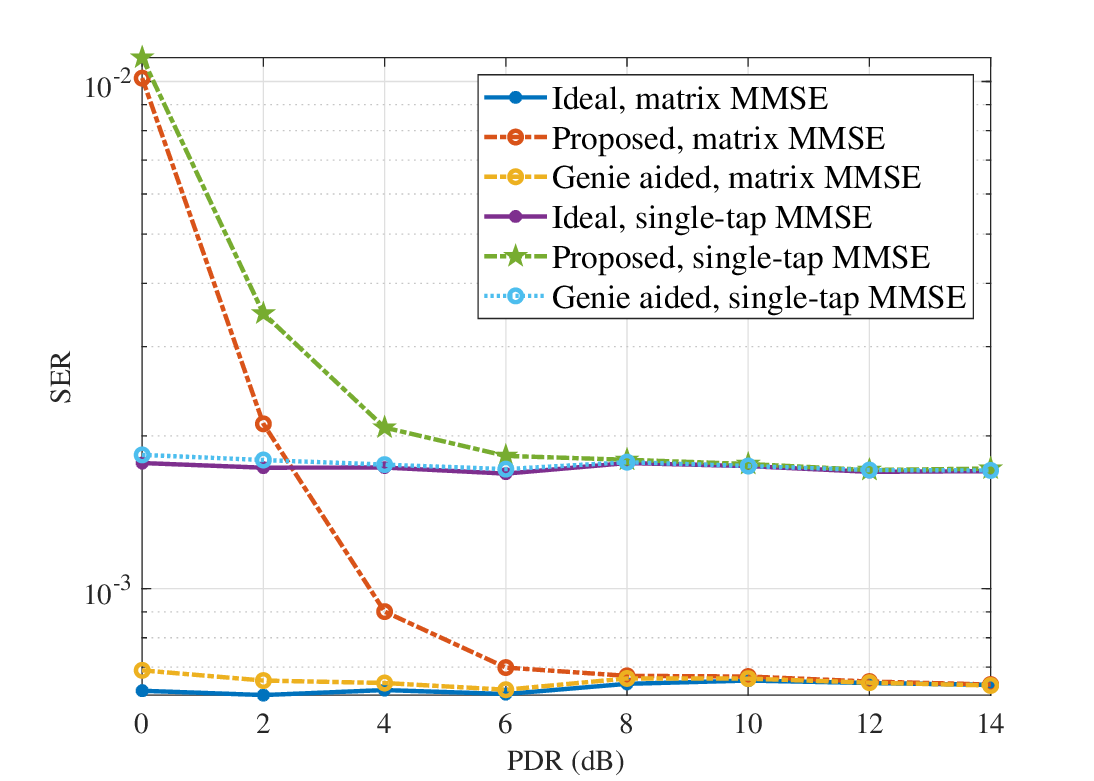}  
    \vspace{-2mm}
    \caption{SER performance with different PDRs.}
    \label{PSR300}\vspace{-6mm}
\end{figure}
Fig. \ref{PSR300} illustrates the SER performance versus PDR at a speed of 300km/h and SNR=30dB. \textcolor{black}{It can be observed that as PDR increases, the SER gradually decreases for both full-matrix and single-tap MMSE equalizers. When the PDR exceeds 8 dB, the SER performance of the proposed CD-IED scheme approaches that of the corresponding Ideal CSI and Genie-aided cases. In fact, the time-domain PP sequence is coherently combined in the DD-domain, whereas interference from data symbols are incoherently combined and averaged across DD grids. Consequently, the proposed scheme requires only a relatively low PDR to achieve performance close to that of ideal CSI and genie-aided schemes, thereby avoiding high PAPR in the time-domain signal.}

\begin{figure}[t]
    \centering
    \vspace{-3mm}
    \includegraphics[width=3in]{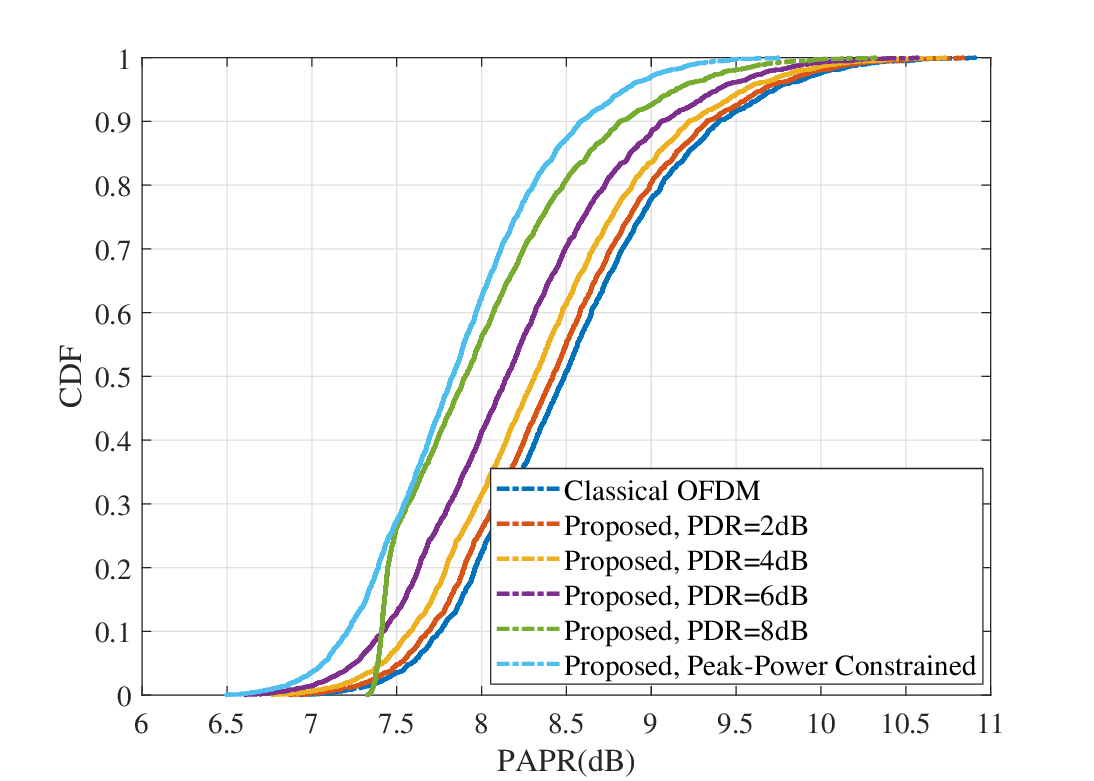}  
    \vspace{-1.5mm}
    \caption{CDF comparison of PAPR performance.}
    \label{PAPR}\vspace{-0.5mm}
\end{figure}
\textcolor{black}{Fig. \ref{PAPR} illustrates the cumulative distribution function (CDF) comparison of the PAPR performance between classical OFDM and the proposed DD-a-OFDM scheme. The PP power design based on the peak-power constraint introduced above is labeled as ‘Peak-Power Constrained’ in the figure. Note that, for a fair comparison between the $M$-subcarrier OFDM system and the proposed DD-a-OFDM structure, where each frame contains $M_{\mathrm{De}}N$ symbols, the CDF is evaluated based on the concatenation of $N$ consecutive OFDM symbols with CP insertion \cite{PAPR}. It can be observed that the proposed DD-a-OFDM scheme achieves lower PAPR levels than classical OFDM under different PDR settings. Notably, the PAPR further decreases as the PDR increases. This is because the increased PP power introduces additional signal energy and improves the overall average transmit power, while the peak power of the transmitted signal does not increase significantly. Moreover, the ‘Peak-Power Constrained’ design improves the average signal power while maintaining the peak power level of the original OFDM signal, thereby achieving the lowest PAPR.  Therefore, the proposed scheme improves the overall PAPR performance, and thus avoids additional design pressure on the power amplifier (PA). This further demonstrates that the proposed DD-a-OFDM scheme maintains compatibility with existing OFDM transmitter architectures and has practical application potential.}

\begin{figure}[t]
    \centering \vspace{-1mm}
    \includegraphics[width=3in]{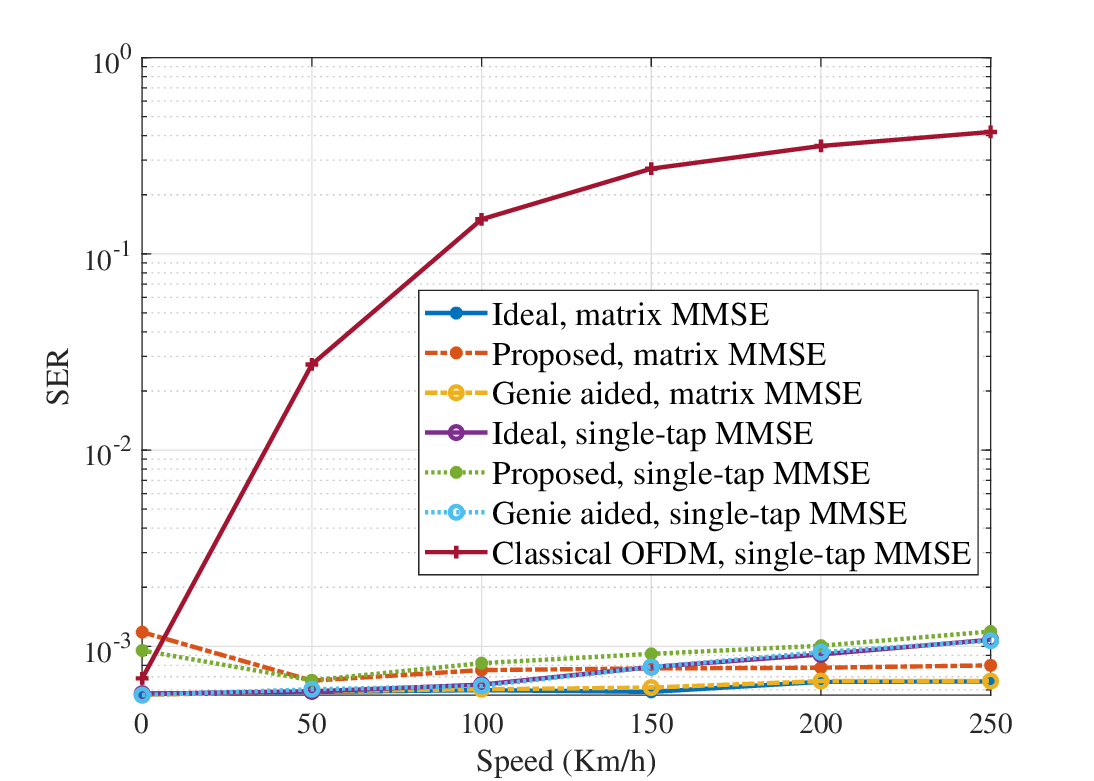}\vspace{-2mm}  
    \caption{SER performance with different speeds in the fractional Doppler case.}
    \label{speed}\vspace{-7mm}
\end{figure}
Fig. \ref{speed} illustrates the SER performance versus the user velocity at SNR=30dB in the fractional Doppler case. \textcolor{black}{It can be observed that the classical OFDM scheme achieves satisfactory performance only under low-mobility or quasi-static channel conditions. As the velocity increases, the rapidly time-varying channel leads to larger CE errors with interpolation-based methods and more severe Doppler-induced ICI, resulting in significant SER degradation. In contrast, the proposed CD-IED algorithm exploits accurate DD-domain parameter estimation to improve TF-domain channel reconstruction and therefore maintains strong robustness against mobility even with a low-complexity single-tap MMSE equalizer. With the full-matrix MMSE equalizer, the reconstructed ICI channel structure can be further exploited to suppress ICI and improve the detection performance.} 

\section{Conclusion}
This paper proposed a DD-a-OFDM transceiver architecture that bridges the complementary strengths of DDMC and OFDM while mitigating their respective limitations. The key innovation lies in the PP design: a single impulse in the DD-domain manifests as a periodic pulse train in the time domain, enabling seamless integration with existing OFDM waveforms without additional interference to data symbols. The proposed PP design preserves the low PAPR of OFDM by spreading pilot power across multiple symbols, significantly reduces pilot overhead compared to conventional TF-domain pilot arrangements, and enables high-resolution delay–Doppler estimation by exploiting the inherent sparsity and stability of the DD-domain channel. Building on this architecture, we developed a novel CD-IED reception algorithm. Since the received PP sequence is coherently combined in the DD-domain, while data symbols are incoherently combined and behave approximately as Gaussian interference, the proposed CD-IED algorithm progressively refines channel estimates and symbol detection by exchanging information between the TF and DD-domains. Simulation results demonstrated that the proposed CD-IED algorithm converges within a few iterations and achieves near-ideal performance. Compared to classical OFDM with TF-domain pilots, the proposed scheme exhibits robust SER performance, while classical OFDM fails due to fast time-varying channel and severe ICI. The full-matrix MMSE equalizer, enabled by the TF-domain channel reconstruction model, effectively suppresses ICI and outperforms single-tap MMSE equalization, particularly in high-mobility scenarios. The proposed DD-a-OFDM framework offers a practical pathway for upgrading existing OFDM-based infrastructure to support high-mobility 6G applications. Future research may explore adaptive PP power allocation, integration with advanced multi-antenna techniques and sequential receiver design.

\bibliographystyle{IEEEtran}
\bibliography{reference}
\end{document}